\documentclass[lettersize,journal]{IEEEtran}
\usepackage{amsmath,amsfonts}
\usepackage{array}
\usepackage[caption=false,font=normalsize,labelfont=sf,textfont=sf]{subfig}
\usepackage{textcomp}
\usepackage{stfloats}
\usepackage{url}
\usepackage{verbatim}
\usepackage{graphicx}
\usepackage{cite}

\usepackage{booktabs}
\usepackage{multirow}
\usepackage{array}
\usepackage[ruled,vlined]{algorithm}  
\usepackage{algpseudocode}

\makeatletter
\newenvironment{breakablealgorithm}
{
		\renewcommand{\ALG@name}{Algorithm}
		\begin{center}
			\refstepcounter{algorithm}
			\hrule height.8pt depth0pt \kern2pt
			\renewcommand{\caption}[2][\relax]{
				{\raggedright\textbf{\ALG@name~\thealgorithm} ##2\par}%
				\ifx\relax##1\relax 
				\addcontentsline{loa}{algorithm}{\protect\numberline{\thealgorithm}##2}%
				\else 
				\addcontentsline{loa}{algorithm}{\protect\numberline{\thealgorithm}##1}%
				\fi
				\kern2pt\hrule\kern2pt
			}
		}{
		\kern2pt\hrule\relax
	\end{center}
}

\begin{document}

\title{Real-Time Bit-Level Encryption of Full High-Definition Video Without Diffusion}

\author{Dong Jiang, Hui-ran Luo, Zi-jian Cui, Xi-jue Zhao, Lin-sheng Huang, Liang-liang Lu
\thanks{This work was supported in part by the National Natural Science Foundation of China under Grant 12274233, in part by the Major Scientific Research Project of Anhui province under Grant KJ2021ZD0005, in part by the Compiled Scientific Research Planning Project of Anhui University under Grant 2022AH040020, and in part by the University Collaborative Innovation Project of Anhui Province under Grant GXXT-2021-091. (\emph{Corresponding authors: Dong Jiang, Lin-sheng Huang, and Liang-liang Lu}.)}
\thanks{Dong Jiang and Lin-sheng Huang are with the School of Internet, Anhui University, Hefei 230039, China, and also with the National Engineering Research Center for Agro-Ecological Big Data Analysis and Application, Anhui University, Hefei, 230601, China (email: jiangd@nju.edu.cn; linsheng0808@ahu.edu.cn).}
\thanks{Hui-ran Luo, Zi-jian Cui, and Xi-jue Zhao are with the School of Internet, Anhui University, Hefei 230039, China.}
\thanks{Liang-liang Lu is with the Key laboratory of Optoelectronic Technology of Jiangsu Province, Nanjing Normal University, Nanjing 210023, China, and also with the National Laboratory of Solid State Microstructures, Nanjing University, Nanjing 210093, China (email: lianglianglu@nju.edu.cn).}
}

\maketitle

\begin{abstract}

Despite the widespread adoption of Shannon's confusion-diffusion architecture in image encryption, the implementation of diffusion to sequentially establish inter-pixel dependencies for attaining plaintext sensitivity constrains algorithmic parallelism, while the execution of multiple rounds of diffusion operations to meet the required sensitivity metrics incurs excessive computational overhead.
Consequently, the pursuit of plaintext sensitivity through diffusion operations is the primary factor limiting the computational efficiency and throughput of video encryption algorithms, rendering them inadequate to meet the demands of real-time encryption for high-resolution video.
To address the performance limitation, this paper proposes a real-time video encryption protocol based on heterogeneous parallel computing, which incorporates the SHA-256 hashes of original frames as input, employs multiple CPU threads to concurrently generate encryption-related data, and deploys numerous GPU threads to simultaneously encrypt pixels.
By leveraging the extreme input sensitivity of the SHA hash, the proposed protocol achieves the required plaintext sensitivity metrics with only a single round of confusion and XOR operations, significantly reducing computational overhead. Furthermore, through eliminating the reliance on diffusion, it realizes the allocation of a dedicated GPU thread for encrypting each pixel within every channel, effectively enhancing algorithm's parallelism.
The experimental results demonstrate that our approach not only exhibits superior statistical properties and robust security but also achieving delay-free bit-level encryption for 1920$\times$1080 resolution (full high definition) video at 30 FPS, with an average encryption time of 25.84 ms on a server equipped with an Intel Xeon Gold 6226R CPU and an NVIDIA GeForce RTX 3090 GPU.
In addition, the proposed protocol is employed to implement a real-time video monitoring system,  enabling delay-free encryption of 640×480 resolution video at 24 FPS on an NVIDIA Jetson Xavier NX equipped with an NVIDIA Carmel ARM CPU and a Volta GPU, thereby demonstrating its feasibility for real-world applications.

\end{abstract}


\newpage
\section{Introduction}
\IEEEPARstart{W}{ith} the rapid advancement of network and multimedia technologies, video has been extensively utilized across various application scenarios, including military command, traffic monitoring, and social networking, consequently creating an substantial demand for secure video transmission and progressively establishing video encryption as a prominent research hotspot \cite{b01}.
To ensure that encrypted video frames attain superior statistical properties and exhibit robust security resilience against potential cryptographic threats, mainstream video encryption algorithms predominantly utilize the Shannon Confusion-Diffusion Architecture (SCDA), a well-established paradigm that has been widely adopted in the field of image encryption \cite{b02}.
In SCDA, confusion systematically permutes pixel positions without modifying their values, thereby disrupting the visual structure of the original frame and reducing statistical correlations among adjacent pixels, 
whereas diffusion strategically alters pixel values and distributes individual pixel influences across the entire frame, thus encrypting the pixels and ensuring plaintext sensitivity \cite{b03}.
However, these operations are inherently time-consuming and commonly require multiple rounds of confusion and diffusion operations to guarantee that the encrypted frames achieve a satisfactory level of statistical properties and security \cite{b04}.

The extended encryption durations, while potentially acceptable for image encryption applications, are insufficient to satisfy the stringent real-time processing demands of video encryption systems, as exceeding the threshold of 1000 milliseconds (ms) divided by the frame rate (Frames Per Second, FPS), potentially resulting in detrimental latency issues.
Therefore, many video encryption algorithms adopt a simplified encryption process, implementing a single round of confusion and diffusion operations for frame encryption, thereby enhancing computational efficiency and encryption speed \cite{b05,b06,b07,b08}.
Although these studies have significantly advanced video encryption technology, their encrypting times, ranging from hundreds to thousands of milliseconds per frame, remain inadequate to satisfy the essential requirements for real-time video encryption applications.
To tackle this challenge, recent works have incorporated parallel computing \cite{b09,b10} and heterogeneous parallel computing \cite{b11} technologies into video encryption, employing CPU thread allocation at the sub-frame level and GPU thread assignment at the pixel level to enable simultaneous execution of confusion and diffusion operations, thereby significantly reducing the average encryption time to below 40 ms, even when a total of  over ten rounds of confusion and diffusion operations are performed on each frame.

\newpage

Despite demonstrating that parallel and heterogeneous parallel computing techniques can effectively enhance encryption speed, these works remain inadequate to meet the demands for real-time encryption of High-Definition (HD) video.
This inadequacy is primarily constrained by two factors: 
First, the diffusion method employed in these works, along with many other approaches \cite{b12,b13,b14}, necessitate the sequential establishment of inter-pixel relationships between original and encrypted counterparts to achieve plaintext sensitivity, thereby ensuring resistance against differential attacks.
The inherently sequential characteristic of diffusion significantly constraints the algorithmic parallelism, limiting the optimal utilization of hardware resources.
Second, the majority of statistical and security metrics can be attained without performing multiple rounds of confusion and diffusion operations. The additional iterations primarily serve to reinforce diffusion effectiveness, thus ensuring compliance with stringent plaintext sensitivity metrics.
The requisite multiple diffusion rounds considerably elevates the computational overhead, compromising encryption speed.
Consequently, the pursuit of plaintext sensitivity through diffusion operations is the primary factor that limits the throughput of video encryption algorithms.

To solve this problem, a real-time video encryption algorithm based on heterogeneous parallel computing is proposed in this paper.
It utilizes the SHA-256 hash of each original frame as input to initialize chaotic systems distributed across multiple CPU threads, which operate in parallel to generate the necessary data for encryption, while allocating numerous GPU threads to concurrently perform bit-level confusion and XOR operations on respective pixels for frame encryption.
Our method ensures that any modification to even a single channel of any pixel in the original frame generates a distinct SHA-256 hash, leading to entirely different data from the CPU threads and resulting in fundamentally divergent encryption outcomes from the GPU threads, thereby achieving plaintext sensitivity without diffusion.
Thus, it can not only allocate a dedicated GPU thread for each channel of every pixel, significantly enhancing the algorithmic parallelism, but also attain required plaintext sensitivity metrics with just a single round of confusion and XOR operations, substantially reducing the computational overhead, ultimately resulting in accelerated encryption speed.
The experimental results demonstrate that the proposed protocol exhibits outstanding statistical properties, provides robust resistance against different types of attacks and channel noise, and enables delay-free full HD (1920$\times$1080 resolution) video encryption at 30 FPS, utilizing a server equipped with an Intel Xeon Gold 6226R CPU and an NVIDIA GeForce RTX 3090 GPU, achieving an average encryption time of 25.84 ms.
It is also utilized to implement a real-time secure video monitoring system that enables delay-free 640×480 video encryption at 24 FPS on an NVIDIA Jetson Xavier NX featuring an NVIDIA Carmel ARM CPU and a Volta GPU, showcasing its high feasibility.
Furthermore, by utilizing the SHA-256 hash of each original frame, our protocol not only establishes a dynamic key space and exhibits resistance to dynamic degradation but also demonstrates enhanced resilience against cropping attacks and channel noise, owing to the absence of diffusion affecting the decryption.

\newpage

\section{Protocol description}
\label{Sec:section2}

The proposed protocol leverages heterogeneous parallel computing, utilizing a main CPU thread to oversee the encryption process, multiple worker CPU threads to concurrently generate different types of data required for encryption, and numerous GPU threads to simultaneously perform bit-level confusion and XOR operations to encrypt the original frame. It consists of three phases: initial condition reconstruction, data generation, and frame encryption.
This section elaborates on the operational workflow of each phase, accompanied by detailed algorithmic descriptions.

\subsection{Initial condition reconstruction}

In the proposed protocol, the Lorenz Hyper Chaotic System (LHCS), defined as follows \cite{b15}, is employed due to its exceptional statistical properties and extensive applications in audio \cite{b16}, image \cite{b17}, and video encryption \cite{b18}.
\begin{equation}
	\left\{
	\begin{array}{lr}
		\begin{split}
			\dot{x}&=\sigma(y-x)+w\\
			\dot{y}&=\rho x-y-xz\\
			\dot{z}&=xy-\beta z\\
			\dot{w}&=-yz + \gamma w\\
		\end{split}
	\end{array}
	\right.
	\label{equ:hyperChaoticSystem}
\end{equation}
where $\sigma$, $\rho$, $\beta$, $\gamma$ are constants, while $x_0$, $y_0$, $z_0$, and $w_0$ are referred to as the initial conditions, with $\gamma$ serving as the control parameter. When $\sigma=10$, $\rho=28$, $\beta=\frac{8}{3}$, and $-1.52\leq\gamma\leq -0.06$, LHCS exhibits a hyperchaotic behavior \cite{b19}.
The main thread $T_m$ utilizes a single LHCS, denoted as LHCS$_m$, whereas each worker thread $T_w^i (i\in\{1, 2,..., n\})$ employs two LHCSs, designated as LHCS$_{w}^{i,1}$ and LHCS$_{w}^{i,2}$, representing the first and second LHCSs of $T_w^i$, respectively.
The LHCS$_m$ is utilized to generate initial conditions necessary to initialize the LHCS$_w$s, which are responsible for producing distinct types of data required for the frame encryption phase.

Unlike many video encryption algorithms that produce encrypted frames by utilizing data derived from an iterative trajectory, the proposed protocol reconstructs the initial conditions for all LHCSs prior to frame encryption, ensuring that each frame is processed using entirely distinct trajectories.
The initial condition reconstruction can be distinguished into two cases, with the user input of $x_0$, $y_0$, $z_0$, $w_0$, and $\gamma$ required for the first original frame, followed by the reconstruction of initial conditions using the SHA-256 hash of the frame.
However, the calculation of the SHA-256 hash for a frame with a resolution of 1920$\times$1080 requires approximately 17 ms, even on an Intel Xeon Gold 6226R CPU, rending it infeasible for real-time video encryption.
Thus, in the proposed protocol, $T_m$ partitions the original frame into multiple sub-frames $f_i (i\in\{1, 2,...,n\})$, with each $T_w^i$ computing the SHA-256 hash $h_i$ for its assigned sub-frame $f_i$. 
Subsequently, $T_m$ calculates the SHA-256 hash $H$ of the frame through the bitwise XOR operation, expressed as $H=h_1\oplus h_2\oplus...\oplus h_n$.

\begin{figure*}[h]
	\centering
	\includegraphics[width=1\textwidth]{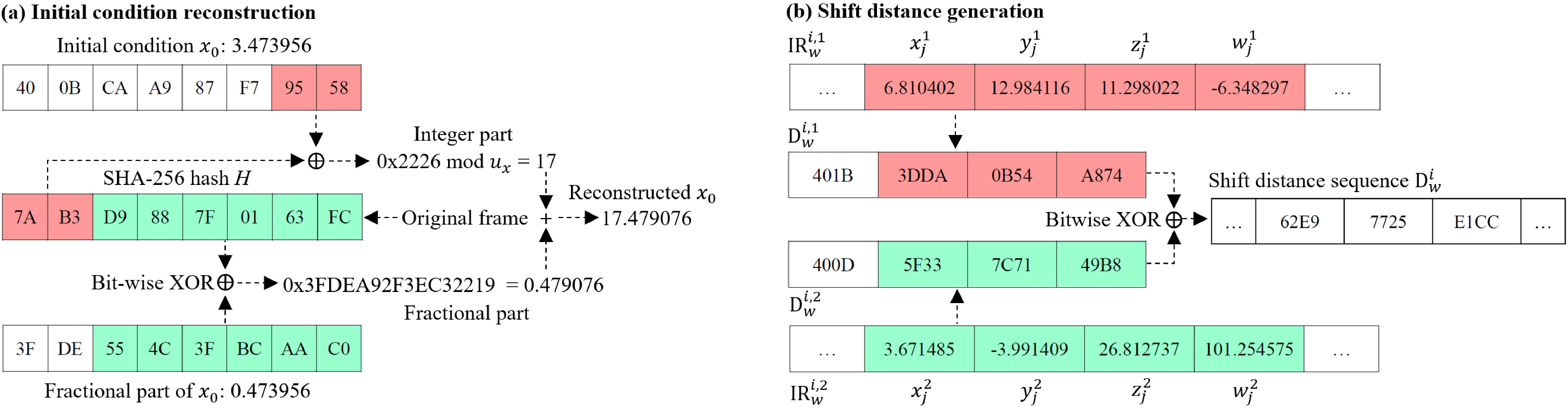}
	\caption{The workflow diagrams of initial condition reconstruction and shift distance generation.}
	\label{fig:figure1}
\end{figure*}

The calculated SHA-256 hash is employed to reconstruct the user-input initial conditions required for initializing the LHCS$_m$, with $x_0$ serving as an example to demonstrate the reconstruction process, as illustrated in Fig. \ref{fig:figure1} (a).
The reconstruction of $x_0$ is achieved through a three-step process: first, generating an integer via a bitwise XOR operation between the two least significant bytes of $x_0$ and two bytes extracted from $H$, followed by a modulo operation constrained by the predefined upper bound $u_x$ of $x_0$; second, producing a decimal through a bitwise XOR operation between the six least significant bytes of the fractional part of $x_0$ and bytes from $H$; and finally, combining these generated integer and decimal values.
The reconstruction of an initial condition requires eight bytes from the SHA-256 hash, with the full 32-byte hash value enabling the reconstruction of all initial conditions $x_0$, $y_0$, $z_0$, and $w_0$, which, in conjunction with the control parameter $\gamma$, are utilized to initialize the LHCS$_m$. Following the initialization, LHCS$_m$ is iterated to produce a set of iteration results $\mathrm{IR}=\{x_1, y_1, z_1, w_1,...,x_{2n}, y_{2n}, z_{2n}, w_{2n}\}$, which are subsequently utilized to initialize all LHCS$_w$s.
For each subsequent frame following the first, the iteration results $x_{2n}$, $y_{2n}$, $z_{2n}$, and $w_{2n}$ from LHCS$_m$ processing the previous frame are reconstructed using the SHA-256 hash of the current frame to generate $x_0$, $y_0$, $z_0$, and $w_0$, which are used to reinitialize LHCS$_m$, followed by reinitializing all LHCS$_w$s using the iteration results from the reinitialized LHCS$_m$.

\subsection{Data generation}

Since the proposed protocol employs the circular shift method to shuffle the pixels and requires bytes for encryption, all $T_w$s utilize their respective LHCS$_w$s to concurrently produce both shift distances and bytes for frame encryption.
For shift distance generation, as illustrated in Fig. \ref{fig:figure1} (b), $T_w^i$ utilizes its respective LHCS$_w$s to produce two sets of iteration results IR$_w^{i,1}=\{..., x_j^1, y_j^1, z_j^1, w_j^1,...\}$ and  IR$_w^{i,2}=\{..., x_j^2, y_j^2, z_j^2, w_j^2,...\}$.  
By extracting six bytes from the mantissa portion\footnote{In IEEE 754 standard, a double-precision floating-point number consists of three parts: a sign bit, an exponent of 11 bits, and a mantissa of 52 bits.} of each iteration result and constructing a shift distance from every two bytes, $T_w^i$ generates two shift distance sequences D$_w^{i,1}=\{d_1^1, d_2^1,...\}$ and  D$_w^{i,2}=\{d_1^2, d_2^2,...\}$.
The shift distance sequence $D_w^i$ for confusion operations can be produced through bitwise XOR operations between D$_w^{i,1}$ and D$_w^{i,2}$, as expressed by $D_w^i= \{d_1^1\oplus d_1^2, d_2^1\oplus d_2^2,...\}$.
The byte generation employs a similar methodology, with the only distinction being that the byte sequence for frame encryption is produced through a bitwise XOR operation applied byte-by-byte to the bytes extracted from two iteration results.

\begin{figure*}[t]
	\centering
	\includegraphics[width=0.96\textwidth]{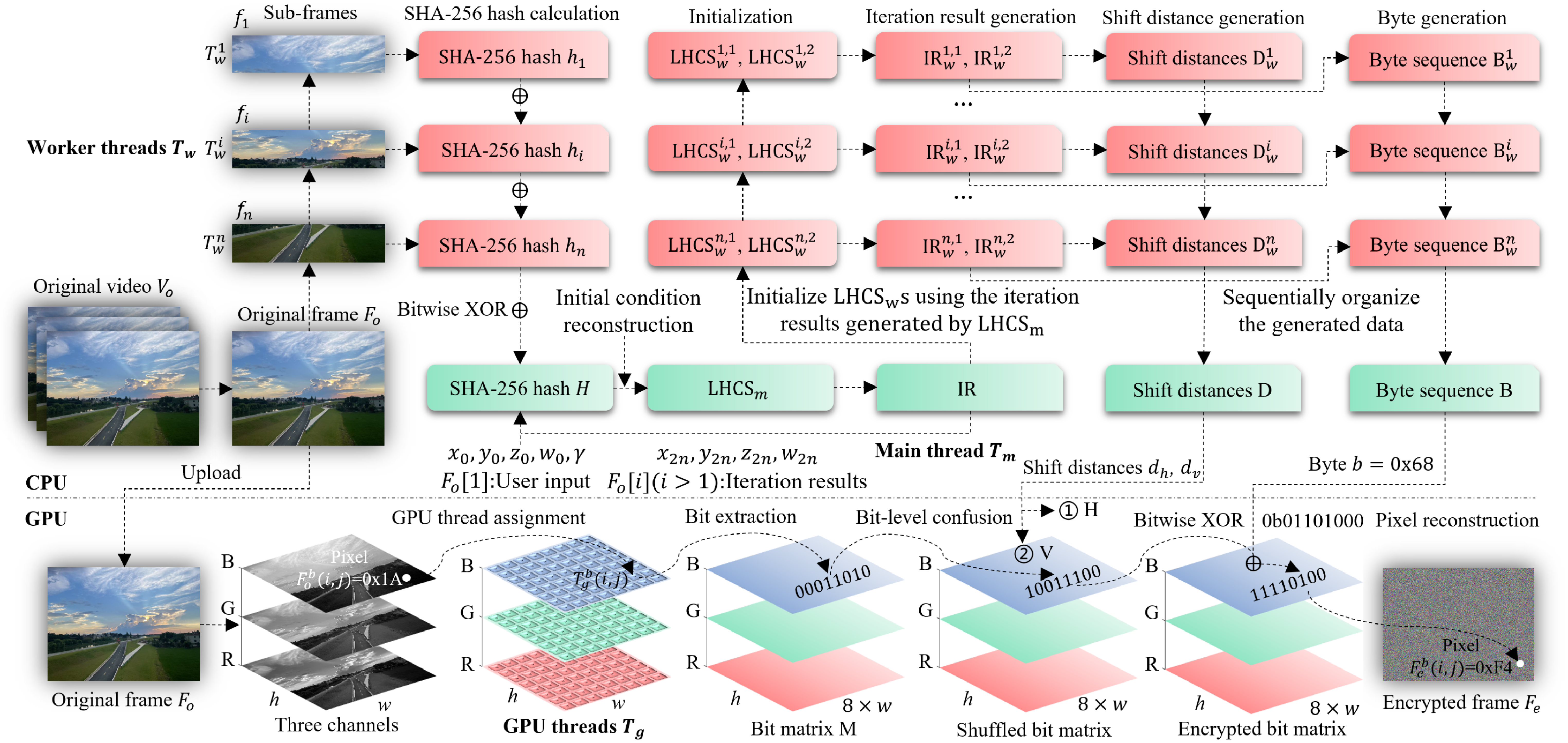}
	\caption{The workflow diagram of the proposed protocol(R: red, G: green, B: blue, H: horizontal, V: vertical, $w$: width, $h$: height.)}
	\label{fig:figure2}
\end{figure*}

\subsection{Frame encryption}

In the frame encryption phase, the generated shift distances $D$ are utilized to perform circular shifts on the frame, thereby shuffling the pixels, while the produced bytes $B$ are employed to apply bitwise XOR operations on the pixels, thus encrypting the frame.
To enhance encryption speed, a dedicated GPU thread $T_{g(i, j)}^c$(where $i\in\{1,..,h\}$ denotes the row index, $j\in\{1,...,w\})$ represents the column index, and $c\in\{r, g, b\}$ indicates the channel) is assigned to each channel of every pixel, enabling the simultaneous execution of confusion and XOR operations.
To encrypt a frame with a resolution of $w\times h$, each $T_g$ initially performs bit-level decomposition of its assigned pixel value, separating it into eight individual bits, thereby generating an expanded bit matrix $M$ of dimension $[8\times w, h]$.
Subsequently, confusion operations involving bidirectional bit-level circular shifting are performed on $M$. For threads sharing the same row and channel, a shift distance $d_h$ retrieved from $D$ enables horizontal circular shifting of their respective 8 bits according to the following equation:
\begin{equation}
	M[i, (j + (d_h~\mathrm{MOD}~8w))~\mathrm{MOD}~8w] = M[i, j], 
\end{equation}
For threads sharing the same column and channel, vertical circular shifting can be applied as follows:
\begin{equation}
	M[(i + (d_v~\mathrm{MOD}~h)~\mathrm{MOD}~h, j] = M[i, j], 
\end{equation}
Finally, each $T_g$ retrieves a byte $b$ from the generated byte sequence $B$, decomposes $b$ into eight individual bits, and performs bit-by-bit XOR operations on its assigned 8-bit pixel value, thereby achieving the encryption of the pixel value.

\subsection{Algorithmic descriptions}

The workflow diagram of the proposed protocol is demonstrated in Fig. \ref{fig:figure2}, accompanied by detailed algorithmic descriptions for the main and worker threads provided in Algo. \ref{alg:algorithm1} and \ref{alg:algorithm2}, respectively, while the algorithms related to the GPU threads are presented in Algo.  \ref{alg:algorithm3} through \ref{alg:algorithm7}.
Given that the decryption process is the inverse transformation of the encryption procedure, its details are not included in this paper. 
However, the complete implementation of the proposed protocol is publicly available in the source code repository at: https://github.com/jiangDongAHU/blfhdve. For further details, please refer to the source code.

\begin{breakablealgorithm}\footnotesize
	\caption{Video encryption algorithm for the main thread $T_m$.}
	\label{alg:algorithm1}
	\begin{algorithmic}[1]
		\Require
		Number of worker threads: $n$; Original video: $V_o$; User inputs: $x_0$, $y_0$, $z_0$, $w_0$,$\gamma$; Frame resolution: width $w$, height: $h$.
		\Ensure
		Encrypted frame: $F_e$.
		\State Create worker threads $\{T_w^i\}_{i=1}^n$;
		\State Allocate matrices $\{M_e^c, M_t^c\}_{c\in\{\mathrm{R, G, B}\}}$ of size $8w\times h$ in GPU memory;
		\While{extract an original frame $F_o$ from $V_o$}
			\State Partition $F_o$ into $n$ sub-frames $\{f_i\}_{i=1}^n$;
			\State Wake up all $T_w$s to compute SHA-256 hash values $\{h_i\}_{i=1}^n$;
			\State Compute SHA-256 hash $H$ of $F_o$ as $H\gets h_1\oplus\cdots\oplus h_n$;
			\If{$F_o$ is the initial frame}
			\State Reconstruct user inputs $x_0$, $y_0$, $z_0$, $w_0$ using $H$;
			\Else
			\State Reconstruct initial conditions using $H$ and $x_{2n}$, $y_{2n}$, $z_{2n}$, $w_{2n}$;
			\EndIf
			\State Initialize LHCS$_m$ using reconstructed $x_0$, $y_0$, $z_0$, $w_0$ along with $\gamma$;
			\State Generate iteration results $\mathrm{IR}=\{x_1, y_1,...,z_{2n}, w_{2n}\}$;
			\State Wake up all $T_w$s to generate shift distances $\mathrm{D}$ and byte sequence $\mathrm{B}$;
			\State Upload $F_o$, $\mathrm{D}$, and $\mathrm{B}$ to GPU memory;	
			\For{$k\gets 3$ \textbf{to} $7$} 
			\State Launch GPU threads $\{T_g^c(i,j)\}_{i=1, j=1}^{h, w}$;
			\State Execute Algo. $k$ concurrently across all GPU threads;
			\State Update $M_t^c\gets M_e^c$;
			\EndFor
			\State Download $F_e$ from the GPU memory;
		\EndWhile
		\State Terminate all worker threads;
	\end{algorithmic}
\end{breakablealgorithm} 

\begin{breakablealgorithm}\footnotesize
	\caption{Data generation algorithm for worker thread $T_w^i$.}
	\label{alg:algorithm2}
	\begin{algorithmic}[1]
		\Require
		Thread index: $i$; Sub-frame: $f_i$; Iteration results generated by $\mathrm{LHCS}_m$: $\mathrm{IR}$.
		\Ensure
		Shift Distances: $\mathrm{D}_w^i$; Bytes: $\mathrm{B}_w^i$.
		\While{True}
		\State Wait to be awakened by $T_m$;
		\State Calculate the SHA-256 hash $h_i$ of $f_i$;
		\State Wait to be awakened by $T_m$;
		\State Fetch iteration results from $\mathrm{IR}$ to initialize $\mathrm{LHCS}_w^{i,1}$ and $\mathrm{LHCS}_w^{i,2}$;
		\State Iterate two $\mathrm{LHCS}_w$s to generate iteration results $\mathrm{IR}_w^{i,1}$ and $\mathrm{IR}_w^{i,2}$;
		\State Generate shift distances $\mathrm{D}_w^{i,1}$ and $\mathrm{D}_w^{i,2}$ using iteration results;
		\State Calculate the final shift distances for confusion $\mathrm{D}_w^i\leftarrow\mathrm{D}_w^{i,1}\oplus\mathrm{D}_w^{i,2}$; 
		\State Generate bytes $\mathrm{B}_w^{i,1}$ and $\mathrm{B}_w^{i,2}$ using iteration results;
		\State Calculate the final bytes for XOR operations $\mathrm{B}_w^i\leftarrow\mathrm{B}_w^{i,1}\oplus\mathrm{B}_w^{i,2}$;
		\EndWhile
	\end{algorithmic}
\end{breakablealgorithm} 

\begin{breakablealgorithm}\footnotesize
	\caption{Bit extraction algorithm for GPU thread $T_g^c(i,j)$}
	\label{alg:algorithm3}
	\begin{algorithmic}[1]
		\Require
		Thread index: $(i, j)$; Channel: $c$; Original frame: $F_o$; Bit matrix: $M_t^c$.
		\Ensure 
		Original bis.
		\State Retrieve the pixel $F_o^c(i,j)$ at coordinate $(i,j)$ from channel $c$ of $F_o$; 
		\For{$k\gets 0$ \textbf{to} $7$} 
			\State temp $\gets$ $F_o^c(i,j)$;
			\State $M_t^c(i,j\times 8 + k)$ $\leftarrow$ temp AND 0x01; {\Comment{Store the bit into $M_t^c$}}
			\State $F_o^c(i,j)\leftarrow F_o^c(i,j) \gg 1$; 
		\EndFor
	\end{algorithmic}
\end{breakablealgorithm}

\begin{breakablealgorithm}\footnotesize
	\caption{Confusion algorithm in the horizontal direction for $T_g^c(i,j)$}
	\label{alg:algorithm4}
	\begin{algorithmic}[1]
		\Require
		Index: $(i, j)$; Channel: $c$; Bit matrices: $M_e^c$, $M_t^c$; Shift distance: $d_h$. 
		\For{$k\gets 0$ \textbf{to} $7$} 
			\State $M_e^c(i, (8j+k + (d_h~\mathrm{MOD}~8w))~\mathrm{MOD}~ 8w)$ $\gets$ $M_t^c(i, 8j+k)$; 
		\EndFor
	\end{algorithmic}
\end{breakablealgorithm}

\begin{breakablealgorithm}\footnotesize
	\caption{Confusion algorithm in the vertical direction for $T_g^c(i,j)$}
	\label{alg:algorithm5}
	\begin{algorithmic}[1]
		\Require
		Index: $(i, j)$; Channel: $c$; Bit matrices: $M_e^c$, $M_t^c$; Shift distance: $d_v$.
		\Ensure 
		Shuffled bits. 
		\For{$k\gets 0$ \textbf{to} $7$} 
		\State $M_e^c((i+(d_v~\mathrm{MOD}~ h))~\mathrm{MOD}~ h, 8j+k)$ $\gets$ $M_t^c(i, 8j+k)$; 
		\EndFor
	\end{algorithmic}
\end{breakablealgorithm}

\begin{breakablealgorithm}\footnotesize
	\caption{XOR operation algorithm for $T_g^c(i,j)$}
	\label{alg:algorithm6}
	\begin{algorithmic}[1]
		\Require
		Index: $(i, j)$; Channel: $c$; Byte extracted from $B$: $b$.
		\Ensure 
		Encrypted bits.  
		\For{$k\gets 0$ \textbf{to} $7$} 
		\State temp$\gets b$ AND 0x01;
		\State $M_e^c(i, 8j +k)\gets M_t^c(i, 8j+k)\oplus$ temp;
		\State $b\gets b\gg 1$; 
		\EndFor
	\end{algorithmic}
\end{breakablealgorithm}

\begin{breakablealgorithm}\footnotesize
	\caption{Pixel reconstruction algorithm for GPU thread $T_g^c(i,j)$.}
	\label{alg:algorithm7}
	\begin{algorithmic}[1]
		\Require
		Index: $(i, j)$; Channel: $c$, Bit matrix: $M_e^c$.
		\Ensure 
		Encrypted pixel $F_e^c(i,j)$. 
		\For{$k\gets 7$ \textbf{to} $0$} 
		\State $F_e^c(i,j)\leftarrow F_e^c(i,j)~\mathrm{OR}~M_e^c(i, (8j+k))$;
		\If{$k\neq 0$}
		\State $F_e^c(i,j)\leftarrow F_e^c(i,j)\ll 1$;
		\EndIf
		\EndFor
	\end{algorithmic}
\end{breakablealgorithm}

\section{Statistical evaluation}
\label{Sec:section3}

An ideal video encryption algorithm should ensure that the encrypted video frames exhibit superior statistical properties, thereby providing robust resistance against various attacks.
Consequently, this section conducts a comprehensive statistical analysis, systematically evaluating the statistical properties of both the original and encrypted frames across three dimensions: uniformity, correlation, and randomness.
To carry out the experiments, six distinct types of original video files, including Akiyo, Coastguard, Hall, Rhinos, Train, and Waterfall, are employed as the experimental dataset.
The proposed protocol is implemented on a workstation equipped with an Intel Xeon Gold 6226R CPU@2.90 GHz, 32 GB of RAM, and an NVIDIA GeForce RTX 3090 graphics card, running the Ubuntu 22.04 operating system with OpenCV 4.5.4 and CUDA 12.4.
For ease of computation, all original video files are converted to MP4 format with a resolution of 512$\times$512 and subsequently encrypted using the implemented cryptosystem.
All original and encrypted frames are stored as images in .tif format, with their statistical properties analyzed utilizing MATLAB R2024b on the Windows 11 operating system.

\subsection{Uniformity evaluation}

\begin{figure}[h]
	\centering
	\includegraphics[width=0.49\textwidth]{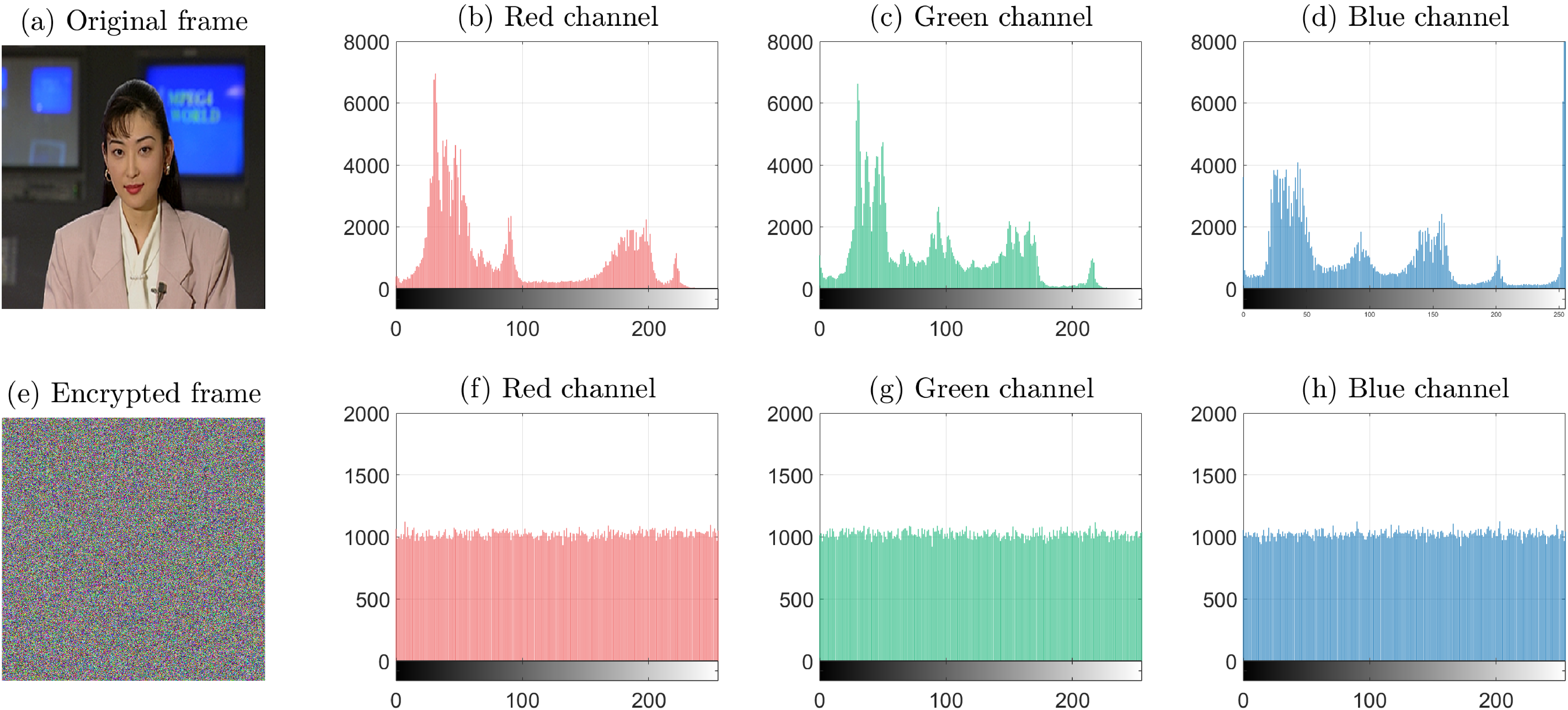}
	\caption{Histograms of the original and encrypted frames, (a) original frame (Akiyo) (b) - (d) histograms of the red, green, and blue channels of the original frame, (e) encrypted frame, (f) - (h) histograms of the red, green, and blue channels of the encrypted frame.}
	\label{fig:figure3}
\end{figure}

A histogram is a graphical representation that depicts the distribution of pixel intensities within a video frame, effectively illustrating the frequency of occurrence for each distinct pixel intensity value \cite{b20}. The implemented cryptosystem should produce encrypted frames characterized by uniformly distributed pixel intensities, thereby ensuring the concealment of the information contained in the original frames.
Therefore, an original frame is selected from the Akiyo video, as shown in Fig. \ref{fig:figure3} (a), with the histograms of its red, green, and blue channels displayed in Fig. \ref{fig:figure3} (b), (c), (d), respectively. 
The corresponding encrypted frame is presented in Fig. \ref{fig:figure3} (e), along with the histograms of its red, green, and blue channels illustrated in Fig. \ref{fig:figure3} (f), (g), (h), respectively.

The variance and $\chi^2$ value of a histogram can be employed to quantitatively measure its uniformity.
The variance of a 256-level intensity histogram for a given frame is calculated using the following equation \cite{b21}:
\begin{equation}
	\mathrm{var}(Z)=\frac{1}{256^2}\sum_{i=0}^{255}\sum_{j=0}^{255}\frac{1}{2}(z_i-z_j)^2,
\end{equation}
where $Z=\{z_0, z_1,..., z_{255}\}$ represents the histogram value vector, with $z_i$ denoting the frequency count of pixels at intensity level $i$. 
$\chi^2$ value of a histogram is defined as \cite{b22}:
\begin{equation}
	\chi^2=\sum_{i=0}^{255}\frac{(z_i-N/256)^2}{N/256},
\end{equation}
where $N$ is the total number of pixels within the frame, $N/256$ indicates the expected frequency of occurrence for each intensity level.
A statistically inverse correlation exists between the variance and the degree of uniformity. At the predetermined significance level $\alpha = 0.05$, the critical $\chi^2$ value with 255 degrees of freedom is determined to be $\chi_{0.05}^2(255) = 293.25$ \cite{b23}.
This indicates that a lower variance in an encrypted frame is preferable, and its $\chi^2$ value must be less than $293.25$.
The variances and $\chi^2$ values of all frames are calculated for each original and encrypted video, with the minimum, maximum, and average variances presented in Tab. \ref{Tab:variances}, and the corresponding $\chi^2$ values demonstrated in Tab. \ref{Tab:chiSquare}. 

\begin{table}[!thbp]
	\centering
	\caption{Variances of the original and encrypted video frames.}
	\begin{tabular}{llccc}
		\toprule
		File Name			& Channel	&Minimum	&Maximum  	  &Average\\
		\midrule
							& Red      	&1549012 	&1588564 	  &1569890  \\
		$V_o$ (Akiyo) 		& Green		&1294011 	&1352300 	  &1321363 \\
							& Blue		&1673799 	&1768554 	  &1716427 \\
		\midrule
							& Red      	&837.733 	&1368.392 	  &1059.840 \\
		$V_e$ (Akiyo) 		& Green		&817.749 	&1431.192 	  &1052.498 \\
							& Blue		&757.357 	&1329.506 	  &1062.486 \\
		\midrule
							& Red      	&428851 	&1032264 	  &716174  \\
		$V_o$ (Coastguard)	& Green		&436678 	&1068929 	  &728323 \\
							& Blue		&526302 	&1094718 	  &812570 \\
		\midrule
							& Red      	&835.584 	&1343.835 	  &1046.121 \\
		$V_e$ (Coastguard) 	& Green		&828.541 	&1298.102 	  &1051.799 \\
							& Blue		&734.455 	&1299.820 	  &1040.726 \\
		\midrule
							& Red      	&1625481 	&1996426 	  &1874129 \\
		$V_o$ (Hall) 		& Green		&1441909 	&1824582 	  &1683879 \\
							& Blue		&1264504 	&1455022 	  &1363269 \\
		\midrule
							& Red      	&800.024 	&1359.624 	  &1068.746 \\
		$V_e$ (Hall) 		& Green		&801.851 	&1335.608 	  &1055.489 \\
							& Blue		&818.525 	&1328.753 	  &1043.324 \\
		\midrule
							& Red      	&1832386 	&15586486 	  &5159395 \\
		$V_o$ (Rhinos)		& Green		&1684264 	&14742950 	  &4497307 \\
							& Blue		&2143286 	&16651285 	  &6392665 \\
		\midrule
							& Red      	&859.647 	&1491.969 	  &1119.121 \\
		$V_e$ (Rhinos) 		& Green		&844.314 	&1438.541 	  &1120.671 \\
							& Blue		&933.263 	&1531.388 	  &1155.267 \\
		\midrule
							& Red      	&1510604    &11445872 	  &3046951 \\
		$V_o$ (Train)		& Green		&1369801 	&6385973 	  &2196022 \\
							& Blue		&2297943 	&9479171 	  &3729807 \\
		\midrule
							& Red      	&818.165 	&1452.596 	  &1096.494 \\
		$V_e$ (Train) 	    & Green		&784.729 	&1338.690 	  &1068.023 \\
							& Blue		&856.737 	&1504.243 	  &1110.727 \\
		\midrule
							& Red      	&731855 	&790814 	  &754425 \\
		$V_o$ (Waterfall)	& Green		&1278808 	&1436426 	  &1347749 \\
							& Blue		&2509192 	&2672742 	  &2582557 \\
		\midrule
							& Red      	&776.361 	&1301.333 	  &1055.060 \\
		$V_e$ (Waterfall) 	& Green		&789.082 	&1327.341 	  &1053.498 \\
							& Blue		&793.451 	&1360.737 	  &1075.248 \\
							
		\midrule
		\multicolumn{5}{l}{$V_o$: original video, $V_e$: encrypted video.}	\\	
		\bottomrule
		\label{Tab:variances}
	\end{tabular}
\end{table}

\begin{table}[!thbp]
	\centering
	\caption{$\chi^2$ values of the original and encrypted video frames.}
	\begin{tabular}{llccc}
		\toprule
		File Name			& Channel	&Minimum	&Maximum  	  &Average\\
		\midrule
							& Red      	&385740 	&395590 	  &390939 \\
		$V_o$ (Akiyo) 		& Green		&322239 	&336754 	  &329050 \\
							& Blue		&416815 	&440411 	  &427431 \\
		\midrule
							& Red      	&208.615 	&340.762 	  &263.925 \\
		$V_e$ (Akiyo) 		& Green		&203.639 	&356.400 	  &262.097 \\
							& Blue		&188.600 	&331.078 	  &264.584 \\
		\midrule
							& Red      	&106794 	&257058 	  &178344  \\
		$V_o$ (Coastguard)	& Green		&108743 	&266188 	  &181370 \\
							& Blue		&131062 	&272610 	  &202349 \\
		\midrule
							& Red      	&208.080 	&334.646 	  &260.509 \\
		$V_e$ (Coastguard) 	& Green		&206.326 	&323.258 	  &261.923 \\
							& Blue		&182.896 	&323.686 	  &259.165 \\
		\midrule
							& Red      	&404783 	&497157 	  &466702 \\
		$V_o$ (Hall) 		& Green		&359069 	&454364 	  &419325 \\
							& Blue		&314891 	&362335 	  &339486 \\
		\midrule
							& Red      	&199.225 	&338.578 	  &266.143 \\
		$V_e$ (Hall) 		& Green		&199.680 	&332.598 	  &262.841 \\
							& Blue		&203.832 	&330.891 	  &259.812 \\
		\midrule
							& Red      	&456307 	&3881400 	  &1284810 \\
		$V_o$ (Rhinos)		& Green		&419421 	&3671340 	  &1119935 \\
							& Blue		&533728 	&4146560 	  &1591924 \\
		\midrule
							& Red      	&214.072 	&371.535 	  &278.687  \\
		$V_e$ (Rhinos) 		& Green		&210.254 	&358.230 	  &279.073 \\
							& Blue		&232.404 	&381.352 	  &287.688 \\
		\midrule
							& Red      	&376176 	&2850290 	  &758762 \\
		$V_o$ (Train)		& Green		&341113 	&1590257 	  &546861 \\
							& Blue		&572242 	&2360536 	  &928809 \\
		\midrule
							& Red      	&203.742 	&361.730 	  &273.053 \\
		$V_e$ (Train) 	    & Green		&195.416 	&333.365 	  &265.963 \\
							& Blue		&213.348 	&374.592 	  &276.597 \\
		\midrule
							& Red      	&182249 	&196931 	  &187870 \\
		$V_o$ (Waterfall)	& Green		&318453 	&357704 	  &335621 \\
							& Blue		&624848 	&665575 	  &643117 \\
		\midrule
							& Red      	&193.332 	&324.062 	  &262.735 \\
		$V_e$ (Waterfall) 	& Green		&196.500 	&330.539 	  &262.346 \\
							& Blue		&197.588 	&338.855 	  &267.762 \\		
		\bottomrule
		\label{Tab:chiSquare}
	\end{tabular}
\end{table}
It is evident that the histograms of all channels of the encrypted frame are smoother than those of the original frame, with the variances and $\chi^2$ values of the encrypted video files exhibiting a notable reduction, and all average $\chi^2$ values remaining below $293.25$, thereby demonstrating the high level of uniformity achieved by the implemented cryptosystem.

\subsection{Correlation evaluation}

\begin{figure}[h]
	\centering
	\includegraphics[width=0.49\textwidth]{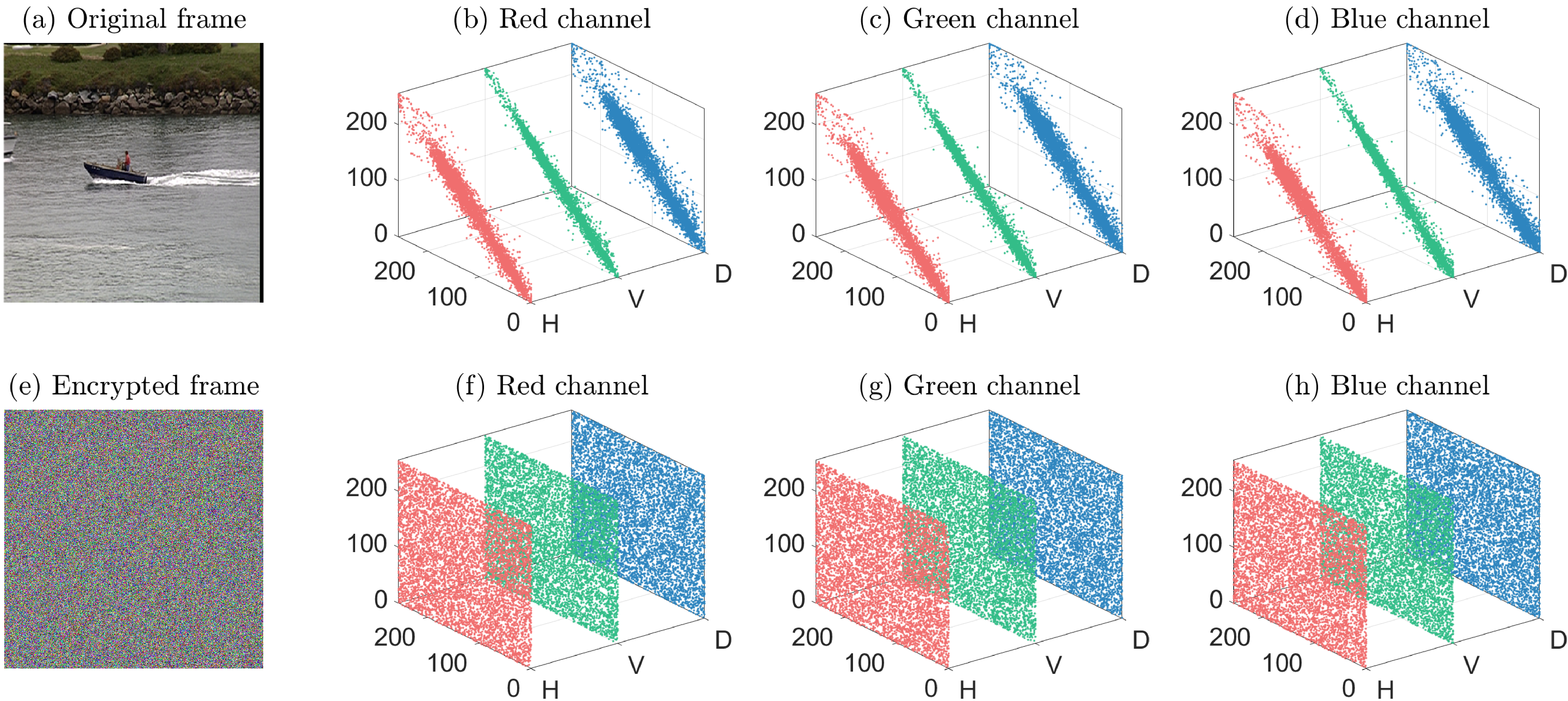}
	\caption{Correlation distributions of adjacent pixel pairs selected from the original and encrypted frames, (a) original frame (Coastguard), (b) - (d) correlation distributions of adjacent pixels selected from the red, green, and blue channels of the original frame along the horizontal, vertical and diagonal directions, (e) encrypted frame, (f) - (h) correlation distributions of adjacent pixels selected from the red, green, and blue channels of the encrypted frame along the horizontal, vertical and diagonal directions.}
	\label{fig:figure4}
\end{figure}

Adjacent pixels within an original frame typically exhibit strong correlations, which must be decorrelated during the encryption process to mitigate the risk of statistical attacks \cite{b24}.
To evaluate the decorrelation performance of the implemented cryptosystem, 6000 pairs of adjacent pixels are randomly selected from each channel of an original frame along different directions.
The original frame is illustrated in Fig. \ref{fig:figure4} (a), while the correlation distributions of the pixel pairs selected from red, green, and blue channels are demonstrated in Fig. \ref{fig:figure4} (b), (c), and (d), respectively. In these figure, the Y-axis represents the values of the selected pixels, the Z-axis corresponds to the values of their adjacent counterparts, and H, V, and D on the X-axis indicate that the pixel pairs are selected along horizontal, vertical, and diagonal directions, respectively.
Similarly, 6000 pairs of adjacent pixels are randomly selected from the encrypted frame, as illustrated in Fig. \ref{fig:figure4} (e), with their correlation distributions for the red, green, and blue channels presented in Fig. \ref{fig:figure4} (f), (g), and (h), respectively.

In addition to correlation distribution, the correlation between adjacent pixel pairs can be quantitatively measured using the correlation coefficient, which is expressed as \cite{b25}:
\begin{equation}
	r_{x,y}=\frac{\mathrm{cov}(x,y)}{\sqrt{D(x)D(y)}},
	\label{Equ:correlationCoefficient}
\end{equation}
where $x$ and $y$ represent adjacent pixel pairs, $\mathrm{cov}(x,y)$ and $D(x)$ can be calculate using the following equations:
\begin{equation}
	\mathrm{cov}(x,y)=\frac{1}{N}\sum_{i=1}^N(x_i-E(x))(y_i-E(y)),
\end{equation}
\begin{equation}
	D(x) = \frac{1}{N}\sum_{i=1}^N(x_i-E(x))^2,
\end{equation}
where $N$ denotes the total number of pixel pairs selected from the frame, and $E(x)$ is defined as:
\begin{equation}
	E(x) = \frac{1}{N}\sum_{i=1}^Nx_i.
\end{equation}

The correlation coefficient ranges from $-1$ to $1$, where a correlation coefficient close to 0 indicates negligible correlation, while a value approaching $\pm 1$ demonstrates a strong correlation.
To quantitatively evaluate the decorrelation performance of the implemented cryptosystem, 10,000 pairs of adjacent pixels are randomly selected from each frame within both the original and encrypted video files.
To avoid the cancellation effect between positive and negative correlation coefficients, the mean of the absolute values of the correlation coefficients is calculated for each video file, with the results presented in Tab. \ref{Tab:correlationCoefficients}.
Evidently, the correlation distributions of all adjacent pixel pairs selected from the original frame are aligned along the diagonal, with the mean absolute correlation coefficients of the original video files close to 1, indicating a strong correlation. 
The correlation distributions of adjacent pixel pairs selected from the encrypted frame are uniformly distributed, with the mean absolute correlation coefficients of the encrypted video files approximately 0, demonstrating a weak correlation.
This clearly highlights the high-level decorrelation capability achieved by the implemented cryptosystem.

\begin{table}[!thbp]
	\centering
	\caption{Mean absolute correlation coefficients of\\adjacent pixel pairs}
	\begin{tabular}{llccc}
		\toprule
		File Name			& Channel	& Horizontal& Vertical	  & Diagonal\\
		\midrule
							& Red      	&0.997044   &0.991953     &0.988869 \\
		$V_o$ (Akiyo) 		& Green		&0.995918   &0.988545     &0.984351 \\
							& Blue		&0.997039   &0.992745     &0.989967 \\
		\midrule
							& Red      	&0.007976   &0.008020  	  &0.007282 \\
		$V_e$ (Akiyo) 		& Green		&0.008356   &0.007697     &0.008246 \\
							& Blue		&0.007876   &0.008046     &0.008175 \\
		\midrule
							& Red      	&0.967330   &0.985433     &0.953370 \\
		$V_o$ (Coastguard)	& Green		&0.969012   &0.986144     &0.955719 \\
							& Blue		&0.974722   &0.988754     &0.963926 \\
		\midrule
							& Red      	&0.008335   &0.008055     &0.008123 \\
		$V_e$ (Coastguard) 	& Green		&0.007766   &0.007960     &0.007950 \\
							& Blue		&0.008037   &0.007638     &0.007445 \\
		\midrule
							& Red      	&0.979658   &0.970518     &0.951983 \\
		$V_o$ (Hall) 		& Green		&0.980364   &0.971818     &0.954248 \\
							& Blue		&0.983515   &0.977858     &0.963319 \\
		\midrule
							& Red      	&0.008971   &0.008458     &0.007930 \\
		$V_e$ (Hall) 		& Green		&0.008037   &0.007861     &0.007906 \\
							& Blue		&0.007849   &0.007739     &0.008121 \\
		\midrule
							& Red      	&0.994612  	&0.994376  	  &0.989562 \\
		$V_o$ (Rhinos)		& Green		&0.994433  	&0.994186  	  &0.989221 \\
							& Blue		&0.993951  	&0.993713  	  &0.988377 \\
		\midrule
							& Red      	&0.008165  	&0.007033  	  &0.006445 \\
		$V_e$ (Rhinos) 		& Green		&0.008355  	&0.007947  	  &0.008863 \\
							& Blue		&0.007620  	&0.007888  	  &0.008514 \\
		\midrule
							& Red      	&0.986818   &0.975006     &0.962045 \\
		$V_o$ (Train)		& Green		&0.987740   &0.976599     &0.964487 \\
							& Blue		&0.980707   &0.963346     &0.944675 \\
		\midrule
							& Red      	&0.008359   &0.008653     &0.008126 \\
		$V_e$ (Train) 	    & Green		&0.007949   &0.007985     &0.007579 \\
							& Blue		&0.009478   &0.008550     &0.008387 \\
		\midrule
							& Red      	&0.978034   &0.971186     &0.954878 \\
		$V_o$ (Waterfall)	& Green		&0.969861   &0.960521     &0.938419 \\
							& Blue		&0.960620   &0.948927     &0.919444 \\
		\midrule
							& Red      	&0.007521   &0.007543     &0.007683 \\
		$V_e$ (Waterfall) 	& Green		&0.007925   &0.008250     &0.008298 \\
							& Blue		&0.008542   &0.008350     &0.008641 \\		
		\bottomrule
		\label{Tab:correlationCoefficients}
	\end{tabular}
\end{table}

\subsection{Randomness evaluation}
In addition to ensuring high uniformity and low correlation, an ideal video encryption algorithm should also guarantee that the encrypted frames exhibit a high level of randomness
Information entropy is a fundamental metric for assessing the level of randomness inherent in a system. For a given information $m$, its information entropy, denoted as $h(m)$, is defined by the following mathematical expression \cite{b26}:
\begin{equation}
	H(m)=\sum_{i=1}^Np(m_i)\mathrm{log}\frac{1}{p(m_i)},
\end{equation}
where $N$ denotes the total number of distinct symbols in the system, and $p(m_i)$ represents the probability of occurrence of the symbol $m_i$.
For a perfectly random information emitting $2^n$ distinct symbols, its information entropy is $n$. Consequently, the theoretical information entropy of a random frame with 256 intensity levels is 8, indicating that an optimal encryption algorithm should generate encrypted frames with an information entropy approaching 8.
The information entropy of all frames are calculated, with the minimum, maximum, and average information entropy for both original and encrypted video files demonstrated in Tab. \ref{Tab:informationEntropy}.
\begin{table}[!thbp]
	\centering
	\caption{Information entropy of the original and encrypted video files}
	\begin{tabular}{llccc}
		\toprule
		File Name			& Channel	&Minimum    &Maximum	  &Average \\
		\midrule
							& Red      	&7.159805 	&7.172551 	  &7.166270 \\
		$V_o$ (Akiyo) 		& Green		&7.237918 	&7.253696 	  &7.245858 \\
							& Blue		&7.233799 	&7.247246 	  &7.240702 \\
		\midrule
							& Red      	&7.999062   &7.999426 	  &7.999274 \\
		$V_e$ (Akiyo) 		& Green		&7.999020 	&7.999440 	  &7.999279 \\
							& Blue		&7.999089 	&7.999481 	  &7.999272 \\
		\midrule
							& Red      	&7.419425 	&7.674670 	  &7.550935 \\
		$V_o$ (Coastguard)	& Green		&7.412202 	&7.664400 	  &7.546955 \\
							& Blue		&7.384193 	&7.610538 	  &7.494974 \\
		\midrule
							& Red      	&7.999081 	&7.999427 	  &7.999283 \\
		$V_e$ (Coastguard) 	& Green		&7.999109 	&7.999433 	  &7.999279 \\
							& Blue		&7.999110 	&7.999496 	  &7.999287 \\
		\midrule
							& Red      	&7.209239 	&7.348232 	  &7.268784 \\
		$V_o$ (Hall) 		& Green		&7.275283 	&7.408671 	  &7.321474 \\
							& Blue		&7.345069 	&7.417197 	  &7.386182 \\
		\midrule
							& Red      	&7.999070 	&7.999451 	  &7.999268 \\
		$V_e$ (Hall) 		& Green		&7.999082 	&7.999451 	  &7.999277 \\
							& Blue		&7.999087 	&7.999440 	  &7.999285 \\
		\midrule
							& Red      	&5.512853 	&7.055801 	  &6.366386 \\
		$V_o$ (Rhinos)		& Green		&5.648948 	&7.008508 	  &6.488776 \\
							& Blue		&5.224948 	&6.932936 	  &6.104077 \\
		\midrule
							& Red      	&7.998978 	&7.999411 	  &7.999233 \\
		$V_e$ (Rhinos) 		& Green		&7.999014 	&7.999423 	  &7.999232 \\
							& Blue		&7.998953 	&7.999360 	  &7.999208 \\
		\midrule
							& Red      	&5.815658 	&7.121010 	  &6.814522 \\
		$V_o$ (Train)		& Green		&6.314428 	&7.178382 	  &6.959167 \\
							& Blue		&5.744660 	&6.779010 	  &6.515665 \\
		\midrule
							& Red      	&7.999008 	&7.999440 	  &7.999248 \\
		$V_e$ (Train) 	    & Green		&7.999083 	&7.999462 	  &7.999268 \\
							& Blue		&7.998969 	&7.999414 	  &7.999239 \\
		\midrule
							& Red      	&7.399047 	&7.429816 	  &7.417382 \\
		$V_o$ (Waterfall)	& Green		&7.004398 	&7.074862 	  &7.042393 \\
							& Blue		&6.528259 	&6.604008 	  &6.562042 \\
		\midrule
							& Red      	&7.999108 	&7.999469 	  &7.999277 \\
		$V_e$ (Waterfall) 	& Green		&7.999090 	&7.999459 	  &7.999278 \\
							& Blue		&7.999066 	&7.999457 	  &7.999263 \\		
		\bottomrule
		\label{Tab:informationEntropy}
	\end{tabular}
\end{table}

Information entropy is commonly utilized to evaluate the overall randomness of frames, whereas local Shannon entropy is employed to quantify the local randomness within those frames. It can be calculated using the following equation \cite{b27}:
\begin{equation}
	\overline{H}_{k,T_B}=\sum_{i=1}^k\frac{H(M_i)}{k},
\end{equation} 
where $M_i(i\in\{1,2,...,k\})$ represents the randomly selected, non-overlapping sub-blocks of a frame, each block comprising $T_B$ pixels, and $H(M_i)$ denotes the information entropy of block $M_i$.
An encrypted frame is considered to pass the local Shannon entropy test if its $\overline{H}_{k,T_B}$ lies within the interval $(h_{left}^*, h_{right}^*)$. According to Ref. \cite{b28}, $k$ and $T_B$ are set to $30$ and $1936$, respectively. For a significance level $\alpha = 0.001$, the idea $\overline{H}_{30,1936}$ is 7.902469317, and the interval $(h_{left}^*, h_{right}^*)$ is $(7.901515798, 7.903422936)$ \cite{b29}.
Thus, thirty non-overlapping sub-blocks are randomly selected from each frame within the original and encrypted video files, with each sub-block comprising $44\times 44$ pixels, totaling 1936 pixels. The $\overline{H}_{30,1936}$ is calculated for each frame, with the minimum, maximum, and average local Shannon entropy presented in Tab. \ref{Tab:localShannonEntropy}.
Evidently, the average information entropy of the encrypted video files all exceed 7.999, and their local Shannon entropy all fall within the interval $(h_{left}^*, h_{right}^*)$, demonstrating that the implemented cryptosystem achieves a high level of both overall and local randomness.

\begin{table}[!thbp]
	\centering
	\caption{Local Shannon entropy $\overline{H}_{30,1936}$ of the original\\and encrypted video files}
	\begin{tabular}{llccc}
		\toprule
		File Name			& Channel	&Minimum    &Maximum	  &Average \\
		\midrule
							& Red      	&4.445855 	&5.373052 	  &4.898197 \\
		$V_o$ (Akiyo) 		& Green		&4.518183 	&5.459264 	  &5.016991 \\
							& Blue		&4.459647 	&5.546943 	  &4.970693 \\
		\midrule
							& Red      	&7.898894 	&7.907350 	  &7.902568 \\
		$V_e$ (Akiyo) 		& Green		&7.898335 	&7.906836 	  &7.902282 \\
							& Blue		&7.898398 	&7.905962 	  &7.902382 \\
		\midrule
							& Red      	&5.745046 	&6.594438 	  &6.216418 \\
		$V_o$ (Coastguard)	& Green		&5.708637 	&6.567717 	  &6.197019 \\
							& Blue		&5.702760 	&6.492304 	  &6.130825 \\
		\midrule
							& Red      	&7.897344 	&7.906669 	  &7.902445 \\
		$V_e$ (Coastguard) 	& Green		&7.896921 	&7.905945 	  &7.902262 \\
							& Blue		&7.898145 	&7.906630 	  &7.902502 \\
		\midrule
							& Red      	&5.063316 	&5.893174 	  &5.493953 \\
		$V_o$ (Hall) 		& Green		&5.018021 	&5.939223 	  &5.458028 \\
							& Blue		&5.257371 	&6.001698 	  &5.634705 \\
		\midrule
							& Red      	&7.898651 	&7.907805 	  &7.902354 \\
		$V_e$ (Hall) 		& Green		&7.897995 	&7.906449 	  &7.902524 \\
							& Blue		&7.897795 	&7.907743 	  &7.902418 \\
		\midrule
							& Red      	&2.448600 	&4.629211 	  &3.553431 \\
		$V_o$ (Rhinos)		& Green		&2.551367 	&4.647639 	  &3.657591 \\
							& Blue		&2.339643 	&4.615171 	  &3.484075 \\
		\midrule
							& Red      	&7.896499 	&7.906353 	  &7.902327 \\
		$V_e$ (Rhinos) 		& Green		&7.898120	&7.906477 	  &7.902420 \\
							& Blue		&7.898046 	&7.906491 	  &7.902306 \\
		\midrule
							& Red      	&3.907096 	&6.165550 	  &5.243988 \\
		$V_o$ (Train)		& Green		&4.034315 	&6.208251 	  &5.377812 \\
							& Blue		&3.868422 	&5.979309 	  &5.159470 \\
		\midrule
							& Red      	&7.897091 	&7.907910 	  &7.902349 \\
		$V_e$ (Train) 	    & Green		&7.897535 	&7.907964 	  &7.902361 \\
							& Blue		&7.898471 	&7.907838 	  &7.902540 \\
		\midrule
							& Red      	&6.423344 	&6.783076 	  &6.614087 \\
		$V_o$ (Waterfall)	& Green		&6.139504 	&6.483053 	  &6.320386 \\
							& Blue		&5.585715 	&5.992534 	  &5.796705 \\
		\midrule
							& Red      	&7.897368 	&7.907045 	  &7.902414 \\
		$V_e$ (Waterfall) 	& Green		&7.897751 	&7.906831 	  &7.902438 \\
							& Blue		&7.897386 	&7.906963 	  &7.902393 \\		
		\bottomrule
		\label{Tab:localShannonEntropy}
	\end{tabular}
\end{table}

\section{Security analysis}
\label{Sec:section4}
A video encryption algorithm must not only generate encrypted frames with superior statistical properties but also exhibit robust resistance against various types of attacks.
Therefore, this section analyzes the resistance of the implemented cryptosystem against several mainstream attack strategies.

\subsection{Resistance to brute-force attacks}
\label{Sec:keySpaceSensitivity}

\begin{figure}[h]
	\centering
	\includegraphics[width=0.49\textwidth]{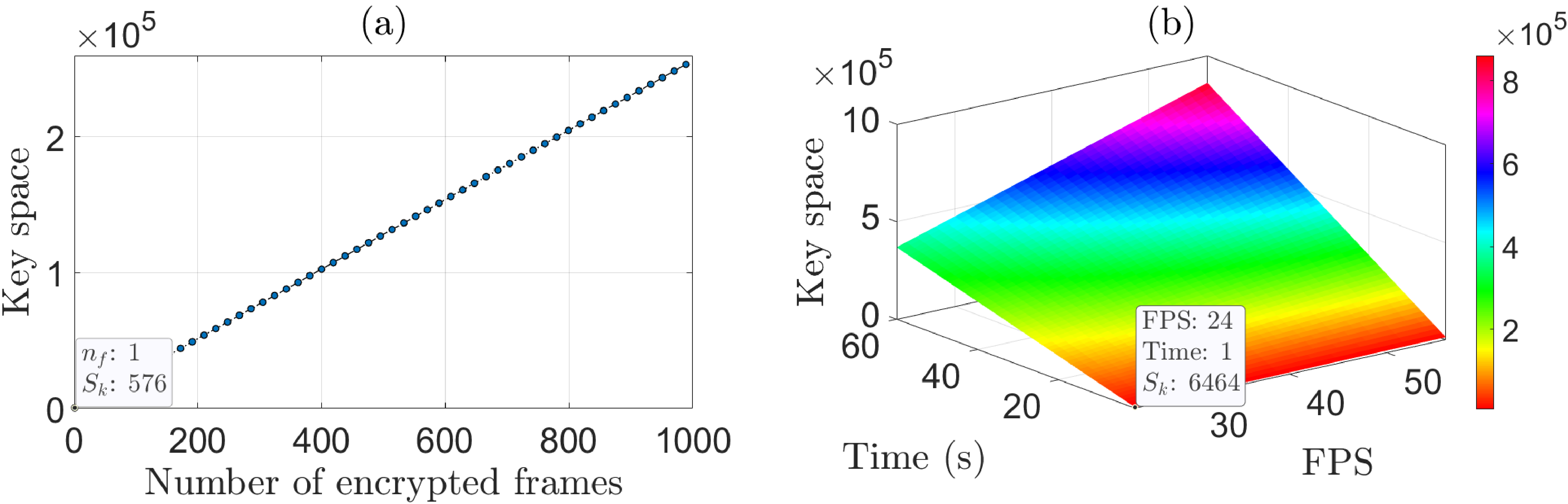}
	\caption{key space of the proposed protocol, (a) relationship between the key space $S_k$ and number of encrypted frames $n_f$, (b) key space $S_k$ as a function of FPS and total encryption time.}
	\label{fig:figure5}
\end{figure}

The key space is a crucial metric for evaluating the security of a cryptosystem. An insufficient key space can render the cryptosystem vulnerable to brute-force attacks, regardless of the robustness of its encryption algorithm \cite{b30}.
Unlike many traditional image or video encryption algorithms that employ fixed key spaces, the proposed protocol reconstructs initial conditions prior to encrypting each frame using the SHA2-256 hash of the original frame.
Clearly, the key consists of the user-input initial conditions $x_0, y_0, z_0, w_0$, control parameter $\gamma$, and all SHA2-256 hashes. The key space of the proposed protocol, denoted as $S_k$, can be defined as follows:
\begin{equation}
	S_k = (5 \times 64) + n_f \times 256,	
\end{equation}
where $n_f$ indicates the number of encrypted frames. The relationship among $S_k$, the FPS of the video, and the total encryption time $t$ (in second) can be expressed as:
\begin{equation}
	S_k = (5 \times 64) + (\mathrm{FPS} \times t) \times 256.	
\end{equation}
The relationship between $S_k$ and $n_f$ is illustrated in Fig. \ref{fig:figure5} (a), while Fig. \ref{fig:figure5} (b) depicts $S_k$ as a function of FPS and $t$.
It is evident that $S_k$ exceeds 550 when encrypting the first frame and surpasses 6400 after one second of encryption. The key space not only exceeds the widely recognized lower bound of $2^{100}$ \cite{b31} but also expands dynamically over time.

\begin{figure}[h]
	\centering
	\includegraphics[width=0.49\textwidth]{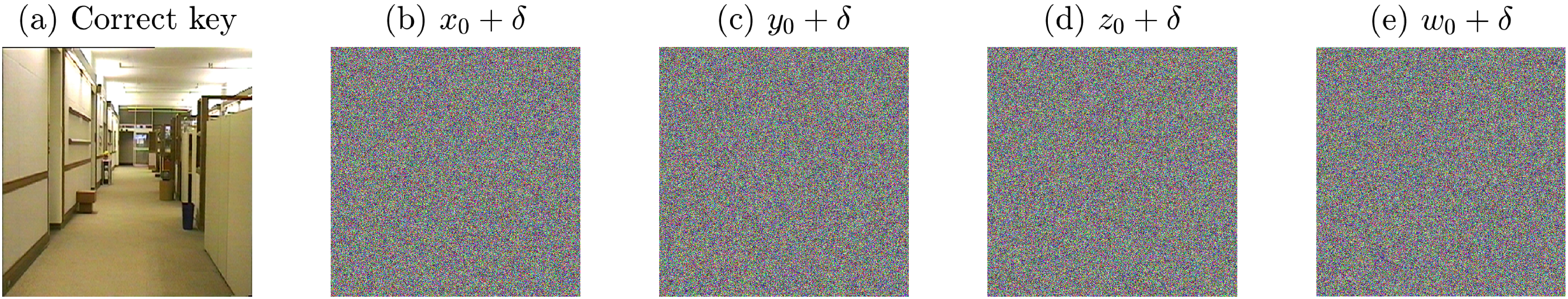}
	\caption{Key sensitivity evaluation, (a) frame (Hall) decrypted with the correct key, (b) frame decrypted with $x_0+\delta$ ($\delta = 1.0\times 10^{-14}$), (c) frame decrypted with $y_0+\delta$, (d) frame decrypted with $z_0+\delta$, (e) frame decrypted with $w_0+\delta$.}
	\label{fig:figure6}
\end{figure}

In addition to possessing a sufficiently large key space, a video encryption algorithm must also demonstrate a high level of key sensitivity \cite{b32}. This ensures that an attacker cannot extract any meaningful information, even when attempting to decrypt the frame using a key that closely resembles the correct one.
To assess the key sensitivity of the implemented cryptosystem, an original frame extracted from the video Hall is encrypted using a randomly selected key. The encrypted frame is then decrypted using the correct key, with the resulting decrypted frame illustrated in Fig. \ref{fig:figure6} (a).
Subsequently, the key is subjected to minimal modification by introducing an increment of $\delta=1.0\times 10^{-14}$ to the initial conditions $x_0$, $y_0$, $z_0$, and $w_0$. The encrypted frame is then decrypted using these slightly altered keys, and the corresponding decrypted frames are depicted in Fig. \ref{fig:figure6} (b) - (e), respectively.
The extensive and dynamically expanding key space, combined with a high degree of key sensitivity, enables the proposed protocol to provide robust resistance against brute-force attacks.

\subsection{Resistance to differential attacks}

\begin{figure}[h]
	\centering
	\includegraphics[width=0.49\textwidth]{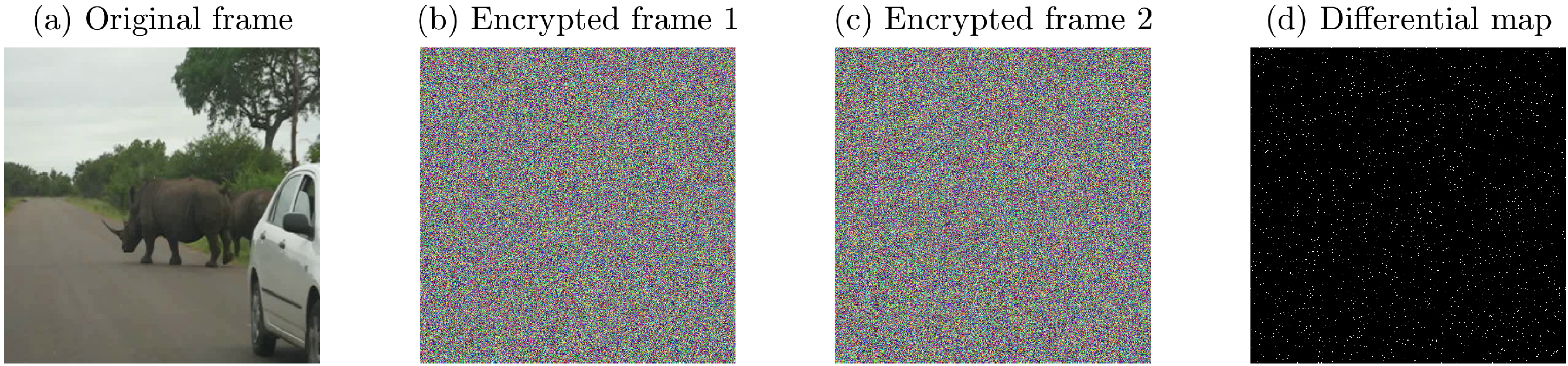}
	\caption{Original frame sensitivity evaluation, (a) original frame (Rhinos), (b) frame obtained by encrypting the original frame using a randomly selected key, (c) frame obtained by encrypting the modified original frame with the same key, (d) differential map of the two encrypted frames.}
	\label{fig:figure7}
\end{figure}

Differential attacks are a category of cryptographic attacks that systematically investigate the correlation between variations in input data and their corresponding transformations in output data during the encryption process \cite{b33}. These attacks leverage statistical analysis of input-output differential patterns to extract information about the encryption key or the internal structure of the cryptographic algorithm.
To counteract such attacks, video encryption algorithms must exhibit a high level of sensitivity to the original frame. Specifically, even a minimal modification of a single pixel in the original frame should result in a completely distinct encrypted frame, despite the use of the same encryption key.
To assess the resistance of the implemented cryptosystem against differential attacks, an original frame from the video Rhinos, as illustrated in Fig. \ref{fig:figure7} (a), is encrypted, resulting in the encrypted frame displayed in Fig. \ref{fig:figure7} (b).
Subsequently, a pixel is randomly selected from the original frame, followed by randomly choosing one of its channel. The pixel value of the selected channel is perturbed by adding a randomly generated increment, resulting in a modified frame, which is encrypted using the implemented cryptosystem with the same key, producing the encrypted frame depicted in Fig. \ref{fig:figure7} (c).
The difference between two encrypted frames is illustrated in Fig. \ref{fig:figure7} (d), where pixels are highlighted in white if any corresponding channel in the two encrypted frames shares identical intensity values.

The differences between two encrypted frames can be quantitatively evaluated through the computation of two metrics: the Number of Pixels Changing Rate (NPCR) and the Unified Averaged Changed Intensity (UACI), defined as follows \cite{b34}:
\begin{equation}
	\mathrm{NPCR} = \sum\limits_{i=1}^w\sum\limits_{j=1}^h\frac{D(i,j)}{{w\times h}}\times 100,
	\label{Equ:NPCR}
\end{equation}
\begin{equation}
	\mathrm{UACI} = \sum\limits_{i=1}^w\sum\limits_{j=1}^{h}\frac{|F_e^1[i,j]-F_e^2[i,j]|}{w\times h\times 255}\times 100,
	\label{Equ:UACI}
\end{equation}
where  $F_e^1$ and $F_e^2$ denote the two encrypted frames, $w$ and $h$ represent frame width and height, $F_e[i,j]$ indicates the pixel value at position $[i, j]$ in frame $F_e$, and $D(i,j)$ is given by:
\begin{equation}
	D(i,j)=
	\left\{
	\begin{array}{lr}
		1, ~F_e^1[i,j] \neq F_e^2[i,j],\\
		0, ~F_e^1[i,j] = F_e^2[i,j].
	\end{array}
	\right.
\end{equation}
Two encrypted frames are considered to successfully pass the NPCR and UACI tests when their NPCR exceeds the critical threshold $N_\rho^*$ and their UACI lies within the critical interval $(U_\rho^-,U_\rho^+)$. Specifically, for encrypted frames with a resolution of 512 $\times$ 512, the critical values are established as $N_\rho^*=99.5893\%$ and $(U_\rho^-,U_\rho^+)=(33.3730\%, 33.5541\%)$ \cite{b35}.

To quantify the resistance of the implemented cryptosystem against differential attacks, all original video files are encrypted using randomly selected keys, generating a set of encrypted video files.
Subsequently, for each original frame, a pixel and a channel are randomly selected, and the intensity value of the selected channel is  modified by adding a randomly generated increment.
The modified original video files are then encrypted using the identical keys, producing a new set of encrypted video files.
Finally, A frame-by-frame calculation of the NPCR and UACI metrics is performed for each corresponding frame pair between the two sets of encrypted video files. The minimum, maximum, and average NPCR and UACI values for all channels across all video files are presented in Tables \ref{Tab:NPCR} and \ref{Tab:UACI}, respectively.
It is evident that all average NPCR values exceed the critical threshold $N_\rho^*$ and all UACI values fall within the critical interval $(U_\rho^-,U_\rho^+)$, indicating that a modification of a single intensity value in any channel of any pixel leads to substantially distinct encrypted frames even when the same key is applied for encryption.

\begin{table}[!thbp]
	\centering
	\caption{Experimental results of NPCR tests}
	\begin{tabular}{llccc}
		\toprule
		File Name			& Channel	&Minimum    &Maximum	  &Average \\
		\midrule
							& Red      	&99.576569 	&99.644852 	  &99.609547 \\
		Akiyo 				& Green		&99.575806 	&99.647522 	  &99.609965 \\
							& Blue		&99.574280 	&99.642563 	  &99.609051 \\
		\midrule
							& Red      	&99.567413  &99.648666 	  &99.609188 \\
		Coastguard			& Green		&99.579620  &99.649048 	  &99.609585 \\
							& Blue		&99.575043  &99.641800 	  &99.609861 \\
		\midrule
							& Red      	&99.574280 	&99.643707 	  &99.609130 \\
		Hall 				& Green		&99.583435 	&99.645615 	  &99.609072 \\
							& Blue		&99.574280 	&99.646378 	  &99.609726 \\
		\midrule
							& Red      	&99.577332 	&99.641418 	  &99.610258 \\
		Rhinos				& Green		&99.570084 	&99.641418 	  &99.610151 \\
							& Blue		&99.560547 	&99.644470 	  &99.610064 \\
		\midrule
							& Red      	&99.570847 	&99.642563 	  &99.608733 \\
		Train				& Green		&99.575043 	&99.647522 	  &99.609879 \\
							& Blue		&99.565887 	&99.646378 	  &99.608375 \\
		\midrule
							& Red      	&99.580002  &99.645233 	  &99.609589 \\
		Waterfall			& Green		&99.564743  &99.652100 	  &99.609184 \\
							& Blue		&99.569321  &99.645615 	  &99.609008 \\	
		\bottomrule
		\label{Tab:NPCR}
	\end{tabular}
\end{table}

\begin{table}[!thbp]
	\centering
	\caption{Experimental results of UACI tests}
	\begin{tabular}{llccc}
		\toprule
		File Name			& Channel	&Minimum    &Maximum	  &Average \\
		\midrule
							& Red      	&33.188037 	&33.733179 	  &33.466065 \\
		Akiyo 				& Green		&33.246472 	&33.702529 	  &33.473039 \\
							& Blue		&33.299864 	&33.641972 	  &33.467086 \\
		\midrule
							& Red      	&33.162613 	&33.680614 	  &33.462089 \\
		Coastguard			& Green		&33.248179  &33.686133 	  &33.464959 \\
							& Blue		&33.307907 	&33.619923 	  &33.468777 \\
		\midrule
							& Red      	&33.039124 	&33.795920 	  &33.458387 \\
		Hall 				& Green		&33.240274 	&33.733515 	  &33.469430 \\
							& Blue		&33.179799 	&33.768077 	  &33.465143 \\
		\midrule
							& Red      	&33.011467 	&33.722690 	  &33.448713 \\
		Rhinos				& Green		&33.216996 	&33.733378 	  &33.481384 \\
							& Blue		&33.144859 	&33.720180 	  &33.474729 \\
		\midrule
							& Red      	&33.030490 	&34.140190 	  &33.475106 \\
		Train				& Green		&33.000192 	&33.911989 	  &33.457124 \\
							& Blue		&32.701047 	&34.107684 	  &33.457344 \\
		\midrule
							& Red      	&33.143939 	&33.751955 	  &33.477481 \\
		Waterfall			& Green		&33.061248 	&33.960244 	  &33.457987 \\
							& Blue		&32.586680 	&34.096925 	  &33.466246 \\	
		\bottomrule
		\label{Tab:UACI}
	\end{tabular}
\end{table}

\subsection{Resistance to cropping attacks and channel noise}
\label{Sec:croppingNoise}

\begin{figure}[h]
	\centering
	\includegraphics[width=0.49\textwidth]{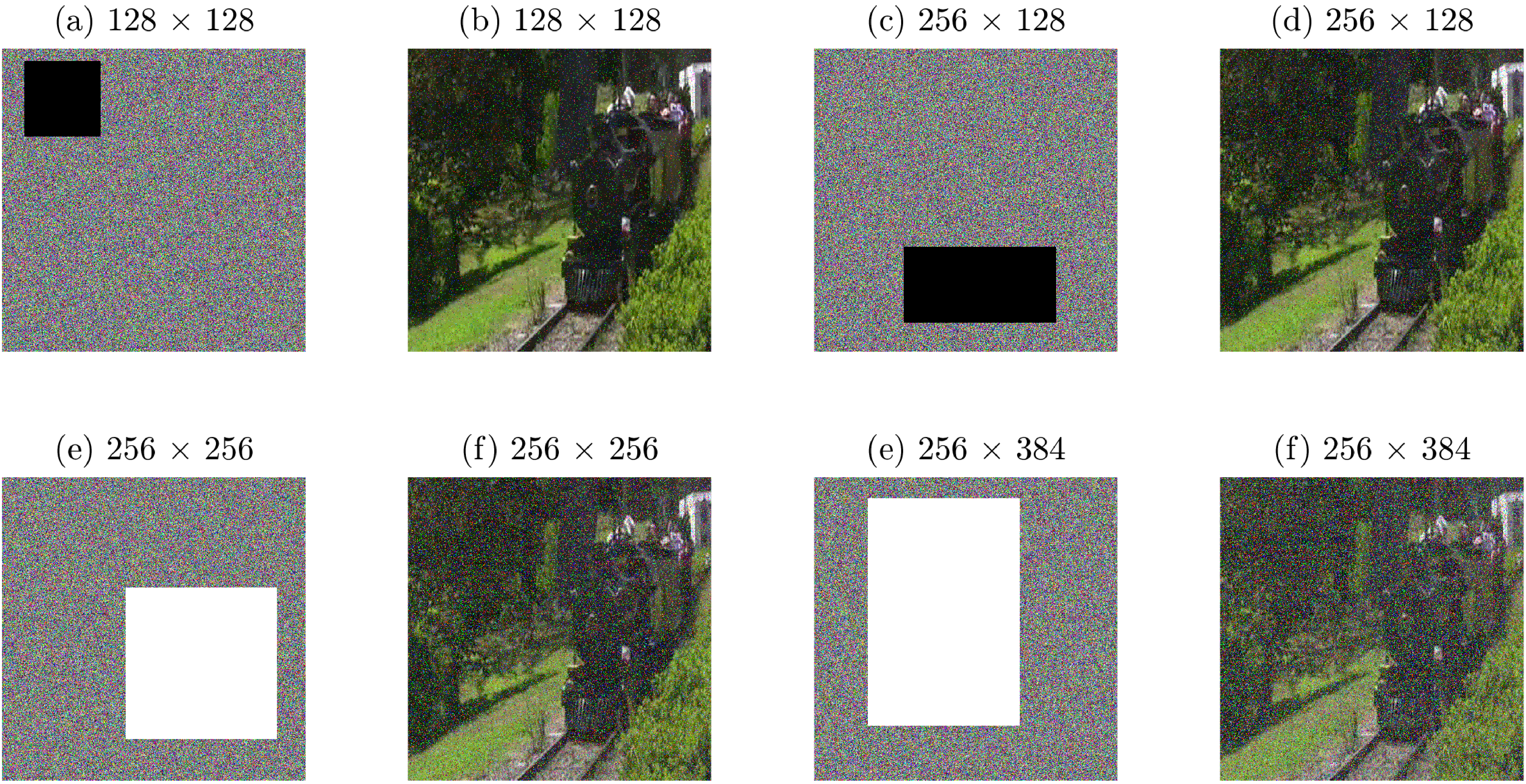}
	\caption{Encrypted frame (Train) after undergoing different degrees of cropping attacks along with their corresponding decryption outcomes. }
	\label{fig:figure8}
\end{figure}

Cropping attacks are a category of cyber threats that involve the unauthorized modification or selective removal of specific regions within encrypted frames, thereby compromising the integrity of these frames and preventing users from accessing accurate information from the original frames \cite{b36}.
Video encryption algorithms must guarantee the restoration of the original frame with superior visual quality, even when the encrypted frame has been subjected to a cropping attack.
To assess the resilience of the implemented cryptosystem against cropping attacks, an original frame is extracted from the video Train and encrypted utilizing a randomly selected key. The resulting encrypted frame is subsequently subjected to different degrees of cropping attacks, with the cropped frame and their corresponding decryption results illustrated in Fig. \ref{fig:figure8}.
Clearly, the contours of the original frame are preserved, even when portions of varying sizes and shapes are cropped from the encrypted frame, including instance where 37.5$\%$ of the frame is removed, thereby highlighting the robustness of the implemented cryptosystem against cropping attacks.

\begin{figure}[h]
	\centering
	\includegraphics[width=0.49\textwidth]{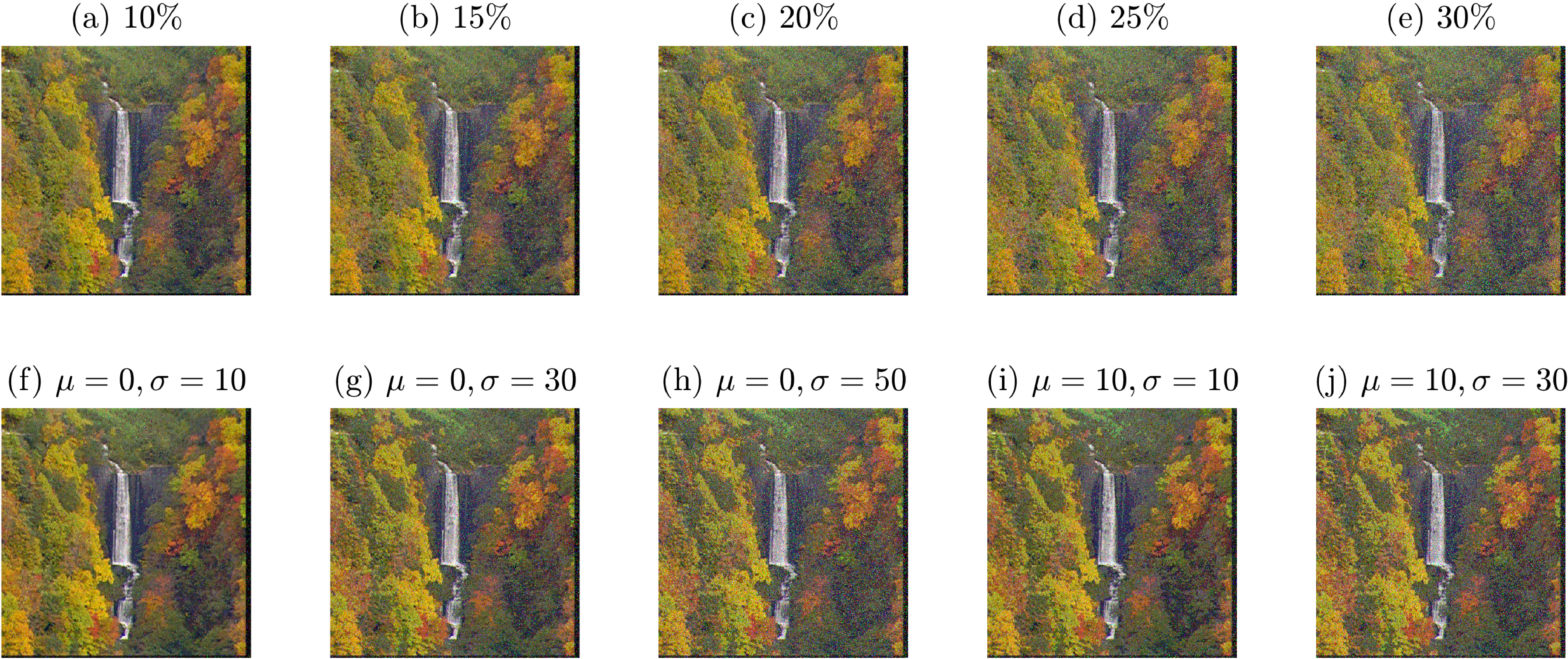}
	\caption{Resistance to channel noise, (a) - (d) decryption results of the encrypted frame (Waterfall) affected by different level of salt-and-pepper noise, (e) - (f) decryption results of the encrypted frame impacted by varying intensities of Gaussian noise. }
	\label{fig:figure9}
\end{figure}

During the transmission process, encrypted frames may be susceptible to various types of noise interference. Therefore, a robust video encryption algorithm should be capable of reconstructing the original frame with high visual fidelity, even in the presence of significant noise disturbances \cite{b37}.
To evaluate the resilience of the implemented cryptosystem against channel noise, an original frame is extracted from the video Waterfall and encrypted to generate an encrypted frame. The resulting frame is then subjected to varying degrees of Salt-and-Pepper noise and Gaussian noise, after which the affected frame is decrypted.
The decrypted frames obtained from the decryption of the encrypted frame containing varying levels of Salt-and-Pepper noise are illustrated in Fig. \ref{fig:figure9} (a) - (e), while those resulting from the decryption of the encrypted frame with different degrees of Gaussian noise are shown in Fig. \ref{fig:figure9} (f) - (j).
It is evident that, even when the encrypted frame is subjected to significant interference from various types of noise, the implemented cryptosystem is still capable of restoring the original frame with high quality.

\section{Application, encryption speed evaluation,\\and comparison}
\label{Sec:section5}

\begin{figure}[h]
	\centering
	\includegraphics[width=0.49\textwidth]{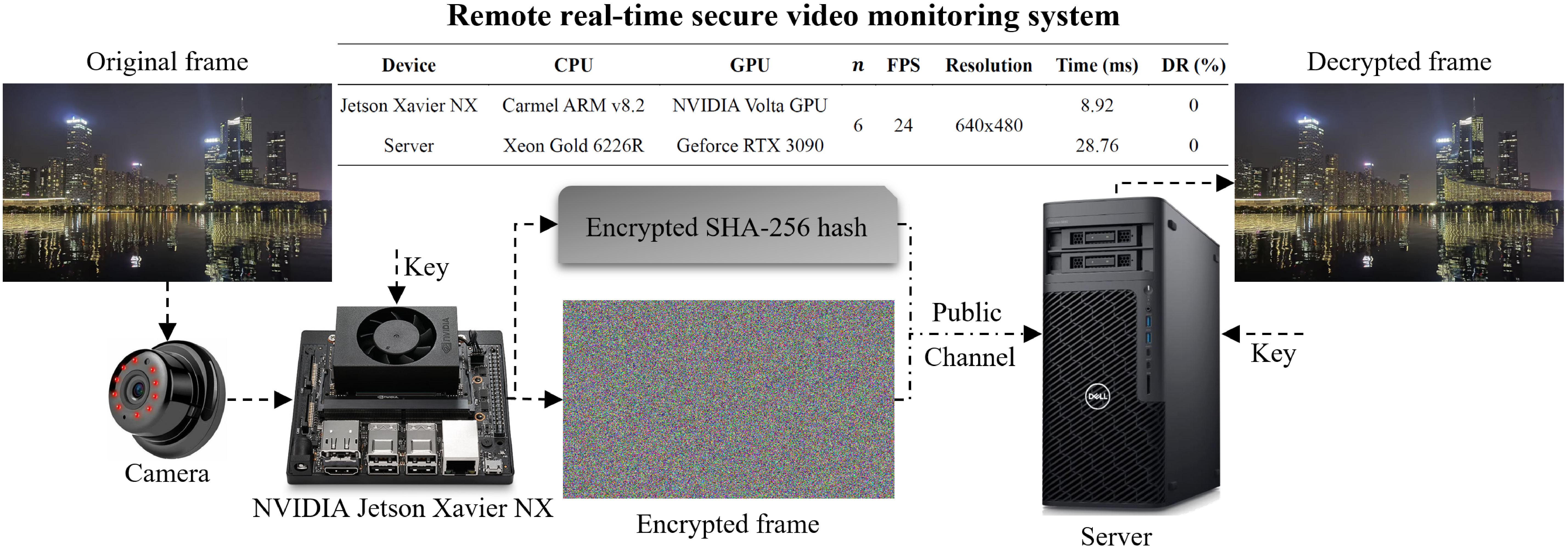}
	\caption{A remote real-time secure video monitoring system deployed using the proposed protocol, $n$: number of worker threads, DR: Delay Rate.}
	\label{fig:figure10}
\end{figure}

Given that video frames may be processed across diverse computing platforms, including servers, personal computers, and even micro-embedded systems, the proposed protocol is employed to implement a remote real-time secure video monitoring system, as illustrated in Fig. \ref{fig:figure10}, to assess its feasibility and practicality.
In the deployed system, original frames are captured and encrypted using an NVIDIA Jetson Xavier NX, which is equipped with a 6-core NVIDIA Carmel ARM V8.2 64-bit CPU and a 384-core NVIDIA Volta GPU. The encrypted SHA-256 hashes of the original frames, along with the encrypted frames, are subsequently transmitted through a public channel to a server featuring an Intel Xeon Gold 6226R CPU and an NVIDIA GeForce RTX 3090 GPU. The server facilitates remote real-time video monitoring by decrypting the received frames.
The captured video is configured with a resolution of 640 $\times$ 480, a frame rate of 24 FPS, and the number of worker threads $n$ for both the server and the embedded system is set to 6 to ensure accurate decryption. 
The experimental results demonstrate that the embedded system achieves delay-free frame encryption with an average time of 28.76 ms, while the server performs delay-free frame decryption with an average time of 8.92 ms.
During the encryption and decryption process, a delay event is recorded when the processing time for a frame exceeds the threshold of 1000 ms divided by the frame rates, approximately 41.67 ms in this instance. The delay rate is determined by dividing the total number of delayed frames by the overall number of frames processed.

\begin{table*}[t]
	\centering
	\caption{Encryption speed evaluation of the proposed protocol with different calculation platforms.}
	\begin{tabular}{llcclcc}
		\toprule
		CPU											   &GPU 					 				 &FPS 		   			   & $n_w$	 		    & Resolution 	 	   		      & AET (ms)&DR($\%$) \\
		\midrule
		\multirow{4}{*}{Inetl Xeon Gold 6226R@2.9 Ghz} &\multirow{4}{*}{NVIDIA Geforce RTX 3090} &\multirow{4}{*}{30}	   &\multirow{4}{*}{32} & 640 $~\times$ 480 $~$(VGAR)	  & 6.36	&0			   \\
													   &										 &				           &					& 720 $~\times$ 480 $~$(SD)	      & 8.95	&0			   \\	
													   &										 &		   				   &					& 1280 $\times$ 720 $~$(HD, 720p) & 14.78   &0			   \\	
													   &										 &				           &					& 1920 $\times$ 1080 (FHD, 1080p) & 25.84   &0		       \\
		\midrule
		\multirow{4}{*}{Inetl Core i7-8700@3.2 GHz}    &\multirow{4}{*}{NVIDIA Geforce GTX 1060} &\multirow{4}{*}{30}      &\multirow{4}{*}{12} & 640  $~\times$ 480 $~$(VGAR)	  & 9.11	&0			   \\
													   &										 &				           &					& 720  $~\times$ 480 $~$(SD)	  & 10.36	&0			   \\	
													   &										 &		   				   &					& 1280 $\times$ 720	$~$(HD, 720p) & 27.89   &0			   \\	
													   &										 &				           &					& 1920 $\times$ 1080 (FHD, 1080p) & 65.09   &100		       \\
		\midrule
		\multirow{4}{*}{NVIDIA Carmel ARM CPU}         &\multirow{4}{*}{NVIDIA Volta GPU} 	     &\multirow{4}{*}{24}      &\multirow{4}{*}{6}  & 320  $~~\times$ 240 			  & 18.48	&0			   \\
													   &										 &				           &					& 360 $~~\times$  240		      & 19.48	&0			   \\	
													   &										 &		   				   &					& 640 $~~\times$ 360			  & 23.62   &0			   \\	
													   &										 &				           &					& 640 $~~\times$ 480 $~$(VGAR)    & 28.26   &0		       \\
		\midrule
		\multicolumn{7}{l}{AET: Average Encryption Time, DR: Delay Rate, VGAR: Video Graphic Array Resolution, SD: Standard Definition, HD: High}\\
		\multicolumn{7}{l}{Definition, FHD: Full High Definition.}\\
		\bottomrule
		\label{Tab:speedEvaluation}
	\end{tabular}
\end{table*}

\begin{table*}[b]
	\centering
	\caption{Comparison of the encryption speed between the proposed protocol and recently published works.}
	\begin{tabular}{lclllllllc}
		\toprule
		Algorithm				   			&YP 	&CA    &Level	&CPU						&GPU				&EM				&Resolution	  	  &AET (ms)	  &RTVE ($\%$) \\
		\midrule
		Ref. \cite{b05} 			&2022   &SC    &Pixel	&Core i7-8750H@2.2 GHz		&--					&C:1,D:1		&352$~\times$288  &260  	  &$\times$\\
		\midrule
		Ref. \cite{b06} 			&2023   &SC   &Block	&Ryzen 9 5950x@3.88 GHz		&--					&C:1,D:1		&352$~\times$192  &231  	  &$\times$\\
		\midrule
		Ref. \cite{b07} 		    &2024   &SC   &Pixel	&Core i5-4120U			    &--					&C:1,D:1		&352$~\times$192  &2652  	  &$\times$\\
		\midrule
		Ref. \cite{b09} 			&2024   &PC	   &DNA		&Xeon Gold 6226R@2.9 Ghz	&--					&C:5,D:3,DNA:4	&512$~\times$512  &34.69  	  &\checkmark\\
		\midrule
		\multirow{2}{*}{Ref. \cite{b10}}  &\multirow{2}{*}{2024} &\multirow{2}{*}{PC} &\multirow{2}{*}{Pixel} &\multirow{2}{*}{Xeon Gold 6226R@2.9 Ghz} &\multirow{2}{*}{--} &\multirow{2}{*}{c:5,d:5} &\multirow{2}{*}{768$~\times$768} &36.56 (PLCM) &\checkmark\\
													&					   &				    &						&										  &				       &						 &								  &36.23 (LASM) &\checkmark\\
		\midrule	
		Ref. \cite{b11}    &2024   &HPC   &Pixel	&Xeon Gold 6226R@2.9 Ghz	&Geforce RTX 3090	&C:7,D:6	    &768$~\times$768  &25.12	  &\checkmark\\
		\midrule
		Ref. \cite{b08} 		&2025   &SC   &Pixel	&Core i7-11390H@3.4 GHz	    &--					&C:1,D:1		&512$~\times$512  &34.69  	  &$\times$\\
		\midrule
		Proposed				   			&-- 	&HPC   &Bit		&Xeon Gold 6226R@2.9 Ghz	&Geforce RTX 3090	&C:1,X:1		&1920$\times$1080 &25.84	  &\checkmark\\
		\midrule
		\multicolumn{10}{l}{YP: Year of Publication, CA: Computational Architecture, EM: Encryption Method, AET: Average Encryption Time, DR: Delay Rate, SC: Serial Comp-}\\
		\multicolumn{10}{l}{uting, PC: Parallel Computing, HPC: Heterogeneous Parallel Computing, C: rounds of Confusion operations, D: rounds of Diffusion operations, X: roun-}\\
		\multicolumn{10}{l}{ds of XOR operations, RTVE: Real-Time Video Encryption, --: not specified.}\\
		\bottomrule
		\label{Tab:comparison}
	\end{tabular}
\end{table*}

The objective of this paper is to enhance the parallelism and reduce the computational load of video encryption while ensuring satisfactory statistical properties and security of the encrypted frames, thereby improving the frame processing speed and enabling real-time video encryption and decryption of higher-resolution videos.
Consequently, a performance assessment of encryption speed is conducted across various computing platforms. The original video, Akiyo, is converted into different resolutions and frame rates and subsequently encrypted. The configurations and experimental results of the computing platforms are presented in Tab \ref{Tab:speedEvaluation}.
To the best of our knowledge, this is the first experimental demonstration of real-time bit-level video encryption, with the server, personal computer, and embedded system achieving delay-free encryption for videos at full high definition (1920 × 1080, commonly referred to as 1080p), high definition (1280 × 720, also known as 720p), and video graphics array (VGA) resolution, respectively, all of which broke the speed records of their respective platforms.
The encryption speed of the proposed protocol is also compared with several recently published works, with the results presented in Tab. \ref{Tab:comparison}.

\section{Discussion}
\label{Sec:section6}

This section provides a discussion on the advantages achieved by the proposed protocol from different perspectives.

\subsection{High sensitivity to the original frame}

\begin{figure}[h]
	\centering
	\includegraphics[width=0.49\textwidth]{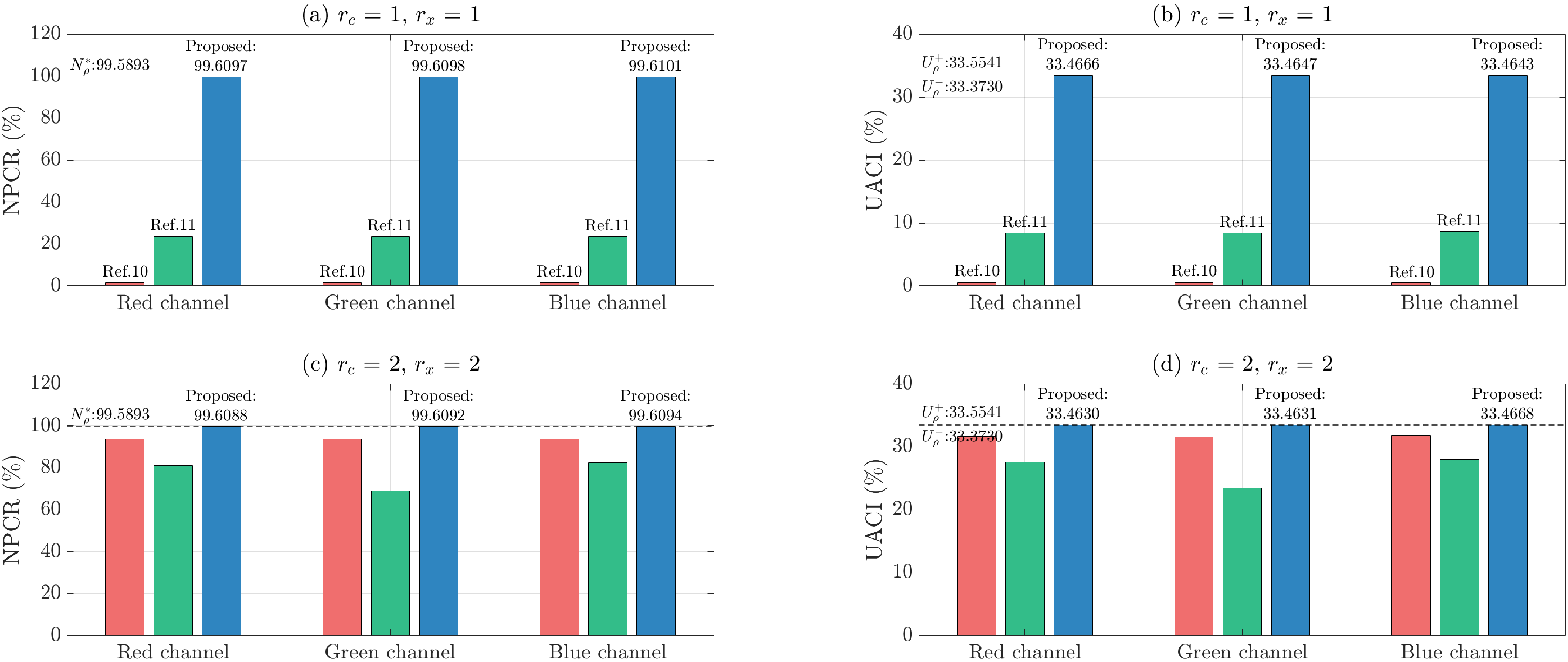}
	\caption{Comparison of original frame sensitivity among different protocols with varying processing rounds ($r_c$: rounds of confusion operations, $r_x$: rounds of diffusion or XOR operations).}
	\label{fig:figure11}
\end{figure}

Among all statistical and security metrics, sensitivity to the original frame is one of the most challenging criteria to satisfy, resulting in numerous image and video encryption algorithms employing multiple rounds of confusion and diffusion operations to provide resistance against differential attacks, which substantially increases encryption time and severely hinders real-time encryption of high-resolution video.
The proposed protocol leverages the extreme sensitivity of the SHA-256 hash function to variations in input, reconstructing the initial conditions using the SHA-256 value of the original frame to ensure that even a single-pixel modification generates a distinct SHA-256 value, resulting in entirely different initial conditions and encryption bytes, thereby enabling the generation of completely different encrypted frames, even when using identical encryption keys.
To validate this advantage, the original video Akiyo is employed to conduct NPCR and UACI tests using the parallel computing-based approach proposed in Ref. \cite{b10}, the heterogeneous parallel computing-based algorithm presented in Ref. \cite{b11}, and the proposed protocol, with different rounds of confusion, diffusion and XOR operations.
The experimental results, as illustrated in Fig. \ref{fig:figure11}, demonstrate that the proposed protocol achieves differential attacks resistance with only one round of confusion and XOR operations, as evidenced by its NPCR value exceeding the critical threshold $N_\rho^*$ and its UACI value falling within the critical interval ($U_\rho^+, U_\rho^-$), while the comparative algorithms fail to pass the tests even with two rounds of confusion and diffusion operations.
The source code for the NPCR and UACI tests of the proposed protocol is publicly accessible in the source code repository at: https://github.com/jiangDongAHU/blfhdve.
 
\begin{figure}[h]
	\centering
	\includegraphics[width=0.49\textwidth]{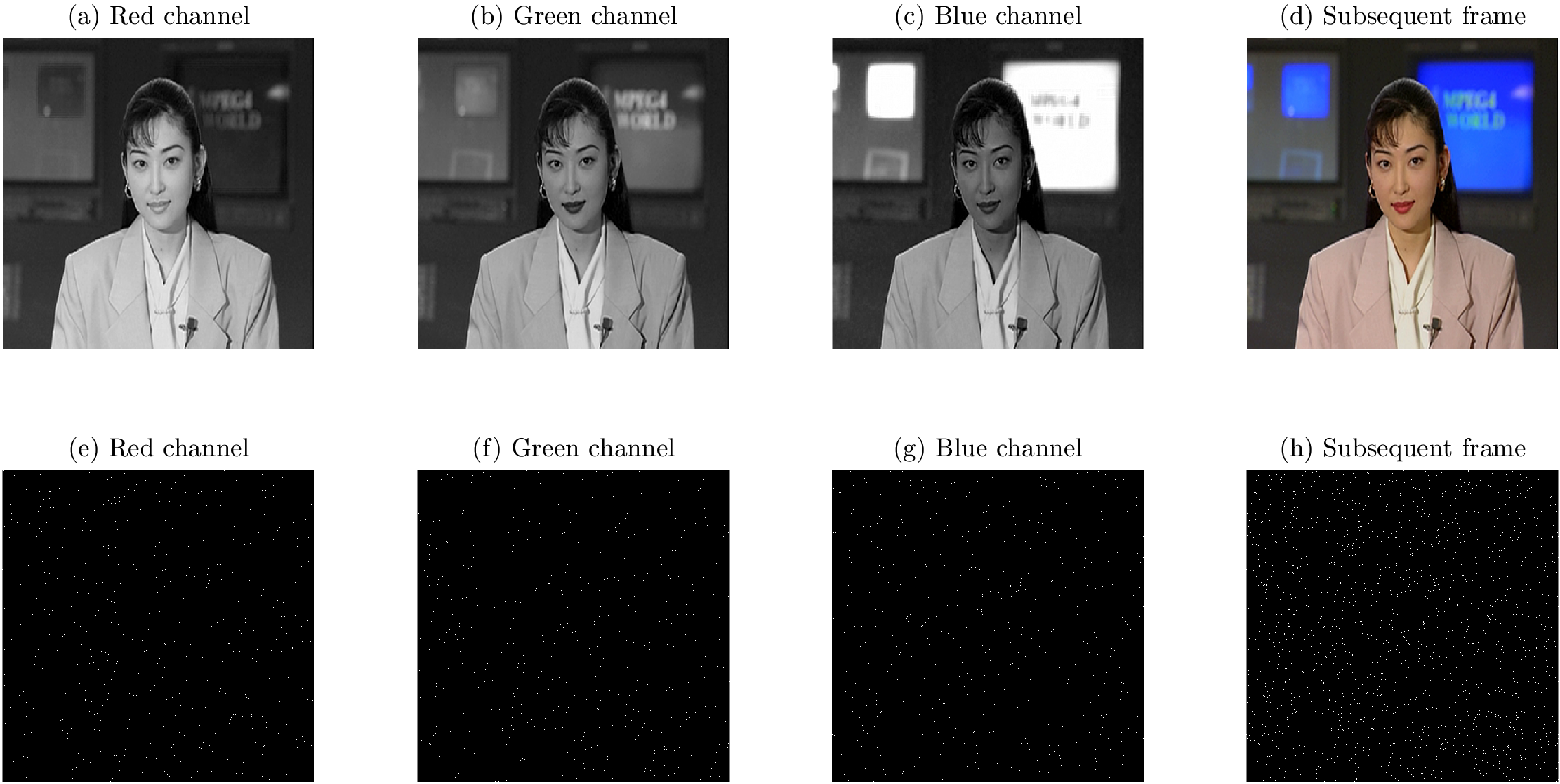}
	\caption{Evaluation of the impact of modifying the pixel value of any channel in the original frame on the encryption results of all channels and subsequent frames, (a) - (c) red, green, and blue channel of the original frame (d) subsequent frame of the original frame (e) - (g) differential maps of the red, green, and blue channels (h) differential map of the subsequent frame (the pixel is marked in white if the pixel values of any channel in the two encrypted frames are identical).}
	\label{fig:figure12}
\end{figure}

Many image and video encryption algorithms process each channel of the original frame, as well as each individual frame, independently. However, as illustrated in Fig. \ref{fig:figure12} (a) - (d), not only do the three channels contain comprehensive information of the original frame, but subsequent frames may also hold substantial amounts of this information.
For these algorithms, altering the value of a particular pixel within a specific channel of the original frame exclusively influences the ciphertext generation of the corresponding channel during the encryption process, without impacting other channels or the encryption of subsequent frames.
In contrast, the proposed protocol ensures that any modification to a pixel value within any channel alters the SHA-256 hash value of the original frame, resulting in a completely distinct encrypted frame, even when the same key is employed.
To validate this advantage, the original frame is encrypted using the implemented cryptosystem to produce an encrypted frame.
Subsequently, a channel is randomly selected from the original frame, a pixel within that channel is chosen and modified by adding a randomly selected increment, after which the modified original frame is encrypted using the same key to generate a new encrypted frame.
The differential maps between the red, green, and blue channels of the two encrypted frames are illustrated in Fig. \ref{fig:figure12} (e) - (g), respectively.
The modification of the original frame significantly impacts the iterative trajectories of the LHCSs, such that while the SHA-256 hash values of subsequent frames remain unchanged, the altered iterative results from the preceding frame lead to the construction of fundamentally different initial conditions, consequently yielding entirely distinct encrypted outputs for all subsequent frames during the encryption process.
As demonstrated by the differential map of the subsequent frame illustrated in Fig. \ref{fig:figure12} (h), although the current frame remains unmodified, its encryption output is entirely altered following the modification of the previous frame. 

\subsection{High algorithmic parallelism}

\begin{figure}[h]
	\centering
	\includegraphics[width=0.36\textwidth]{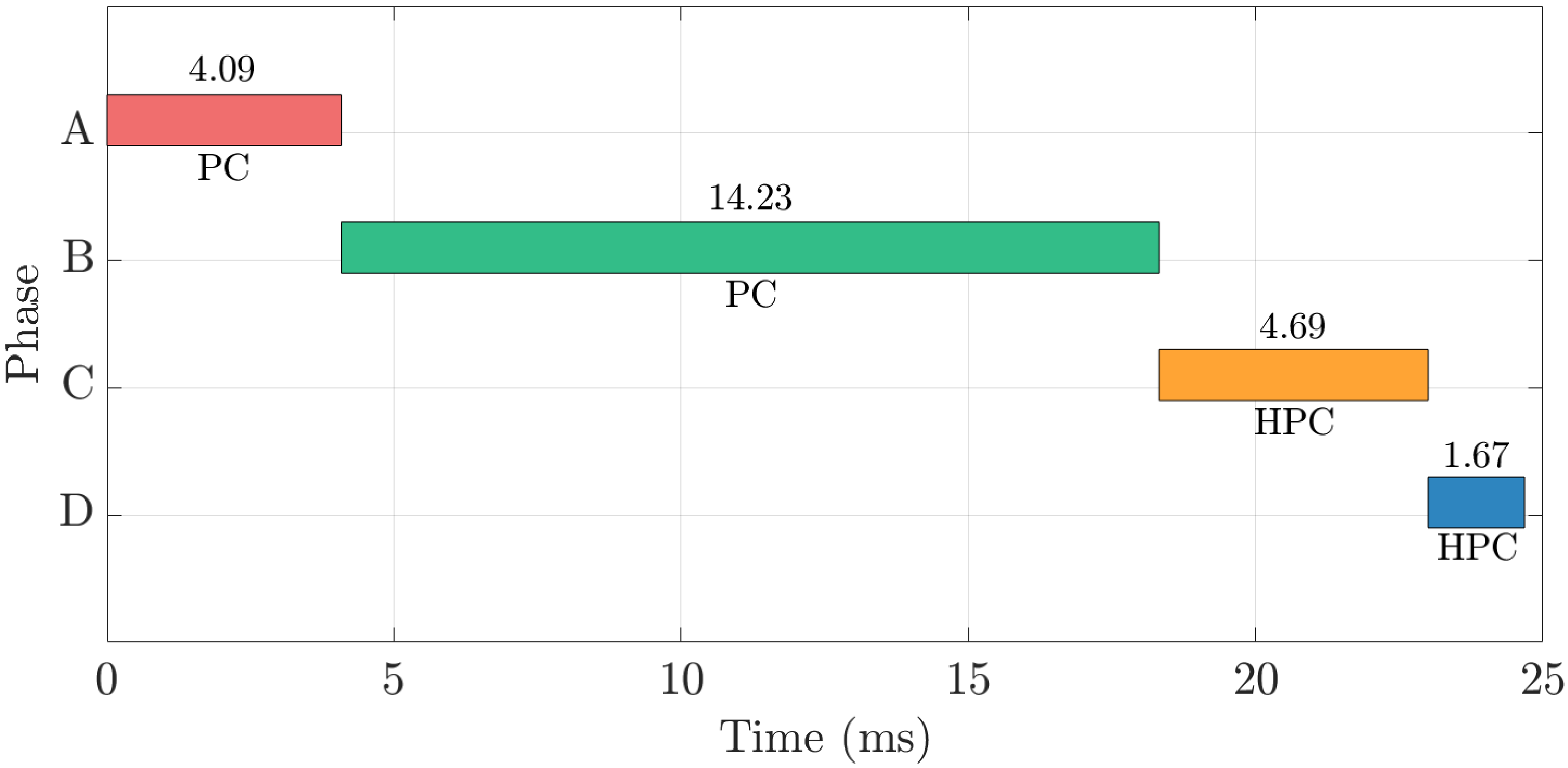}
	\caption{Time consumption of each phase of the proposed protocol, phase A: initial condition reconstruction, phase B: data generation: phase C: bit-level confusion operations, phase D: bit-level XOR operations, PC: Parallel Computing, HPC: Heterogeneous Parallel Computing.}
	\label{fig:figure13}
\end{figure}

The objective of this paper is not only to minimize the computational cost associated with video encryption while ensuring that the encrypted frames exhibit satisfactory statistical properties and security, but also to maximize the parallelism of each phase in the encryption process, thereby enhancing computational efficiency and encryption speed, ultimately achieving real-time encryption of high-resolution video.
As outlined in the protocol description section, the proposed protocol consists of phases for SHA-256 hash calculation, shift distance and byte generation, bit-level confusion operations, and bit-level XOR operations.
The first two phases leverage parallel computing techniques, employing multiple CPU threads to concurrently compute the SHA-256 hash value of the original frame and generate the shift distances and bytes necessary for confusion and XOR operations, while the latter two phases utilize heterogeneous parallel computing techniques, assigning a GPU thread to each channel of every pixel to enable concurrent execution of bit-level confusion and XOR operations.
To assess the time consumption of each phase in the encryption process, the original video Akiyo at a resolution of 1920$\times$1080 is encrypted using the implemented cryptosystem, with the average duration for each phase measured and presented in Fig. \ref{fig:figure13}.
It is evident that the optimization of each stage of encryption through parallel and heterogeneous parallel computing techniques enables the completion of the SHA-256 hash calculation, bit-level confusion, and XOR operations within 5 ms, with the bit-level XOR operation on the entire frame using the generated bytes requiring only 1.67 ms, thereby highlighting the significant speed advantage of the proposed protocol over traditional diffusion-based algorithms.
For the shift distance and byte generation phase, two LHCSs are employed to construct the shift distances and bytes to ensure system security, a process that involves numerous iterations, type conversions, and XOR operations, along with uploading the generated sequences from memory to graphics memory, resulting in an average processing time of 14.23 ms.
However, the experimental results demonstrate that the temporal bottleneck inherent in traditional video encryption algorithms, predominantly attributed to the prolonged computational duration of confusion and diffusion operations, is successfully addressed.

\subsection{Mitigating the effects of dynamical degradation}

\begin{figure}[h]
	\centering
	\includegraphics[width=0.49\textwidth]{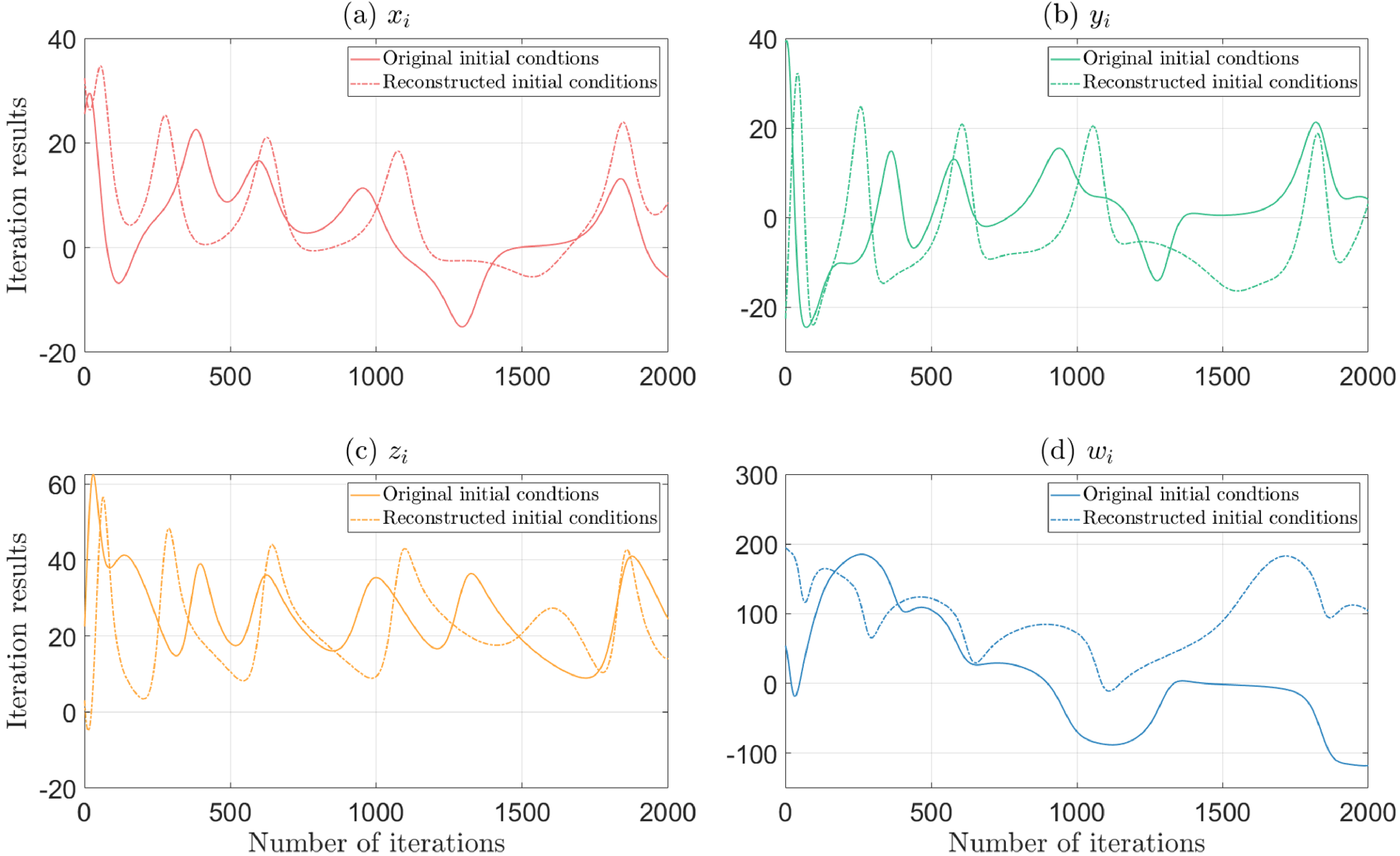}
	\caption{Iteration trajectories of the LHCS with the original and reconstructed initial conditions.}
	\label{fig:figure14}
\end{figure}

When chaotic systems are implemented on finite precision digital devices, their dynamical behaviors frequently diverge markedly from those of the original continuous versions, resulting in cyclic iterative trajectories and degradation of dynamical properties \cite{b38}. 
In the proposed protocol, following the encryption of each frame, the initial conditions of the LHCS in the main thread are reconstructed using the SHA2-256 hash of the subsequent original frame.
The iteration trajectories of LHCS generated from randomly selected initial conditions, as well as those initiated with initial conditions reconstructed using SHA-256 hash value of an original frame, are presented in Fig. \ref{fig:figure14}.
Clearly, the iteration outcomes are sifted to completely different trajectories following the reconstruction of the initial conditions.
The switching of LHCS iterative trajectories in the main thread produces entirely different parameters $P_m$, resulting the shifting of iterative trajectories of LHCSs in the worker threads.
Consequently, all LHCSs in the main thread and the byte generation threads are reinitialized after encrypting a frame, thereby enhancing the effectiveness of mitigating the impact of dynamic degradation compared to traditional methods of perturbing iterative trajectories \cite{b39}.

To sum up, the proposed protocol achieves the following advantages:

\begin{itemize}
	\item[$(1)$] By leveraging the extreme sensitivity of the SHA-256 hash function to input variations, the proposed protocol establishes robust resistance against differential attacks with reduced computational load, while ensuring that any change in the pixel value of any channel affects encryption results of all channels and subsequent frames, thereby achieving high sensitivity to the original frame.
	\item[$(2)$] By eliminating the necessity of diffusion operations to establish the pixel-wise relationship between the original and encrypted pixels, the proposed protocol enables the allocation of individual GPU threads for the parallel encryption of each pixel, simultaneously optimizing other stages of encryption through parallel and heterogeneous computing techniques, thereby significantly enhancing both the algorithmic parallelism and encryption speed.
	\item[$(3)$] By reconstructing the initial conditions using the SHA-256 hash value of each original frame, the proposed protocol not only shifts the iteration trajectories of LHCSs to entirely different paths, thereby mitigating the impacts of dynamic degradation, but also, as discussed in Section \ref{Sec:keySpaceSensitivity}, enables the key space to dynamically expand with the number of processed frames, providing enhanced resistance to brute-force attacks compared to traditional algorithms that rely on a fixed key space. 
	\item[$(4)$] As evidenced in Sec. \ref{Sec:croppingNoise}, the absence of diffusion operations ensures that alterations to a pixel value within the encrypted frame, caused by cropping attacks or channel noise, do not affect the values of other pixels during decryption, thereby providing significant resistance against both cropping attacks and channel noise.
\end{itemize}

\section{Conclusion}
\label{Sec:section7}

To mitigate the computational inefficiencies inherent in diffusion operations, this paper presents a real-time video encryption protocol, leveraging heterogeneous parallel computing.
It incorporates SHA-256 hashes of the original frames as input, utilizes multiple CPU threads to concurrently generate the data necessary for frame encryption, and assigns a dedicated GPU thread to each pixel within every channel to simultaneously perform confusion and XOR operations for pixel encryption.
The statistical evaluation and security analysis demonstrate that our approach exhibits superior statistical properties and provides robust security against different types of attacks.
By utilizing the exceptional sensitivity of SHA hashing instead of relying on multiple rounds of diffusion operations, it achieves high parallelism and low computational overhead, resulting in enhanced encryption speed.
Benchmark results confirm delay-free bit-level encryption at full HD (1080p), HD (720p) and VGA resolutions are achieved across server, desktop, and, embedded implementations, respectively.
The proposed protocol is also successfully implemented in a remote real-time secure video monitoring system, thereby empirically demonstrating both its technical viability and readiness for real-world adoption.
Furthermore, by leveraging SHA-256 hashes of the original frames to eliminate the need for diffusion operations, this approach mitigates the effects of dynamic degradation, establishes a dynamic key space, and provides enhanced resistance against cropping attacks and channel noise.

\vfill
\end{document}